# Cosmic metallicity evolution of Active Galactic Nuclei: Implications for optical diagnostic diagrams

Oli L. Dors[1]⋆, M. V. Cardaci[2,3], G. F. Hägele[2,3], G. S. Ilha[1,6], C. B. Oliveira[1], R. A. Riffel[4], R. Riffel[5], A. C. Krabbe[6]

[1]*Universidade do Vale do Paraíba, Av. Shishima Hifumi, 2911, Cep 12244-000, São José dos Campos, SP, Brazil*
[2] *Facultad de Ciencias Astronómicas y Geofísicas, Universidad Nacional de La Plata, Paseo del Bosque s/n, 1900 La Plata, Argentina.*
[3] *Instituto de Astrofísica de La Plata (CONICET-UNLP), La Plata, Avenida Centenario (Paseo del Bosque) S/N, B1900FWA, Argentina*
[4]*Departamento de Física, Centro de Ciências Naturais e Exatas, Universidade Federal de Santa Maria, 97105-900 Santa Maria, RS, Brazil*
[5]*Instituto de Física, Universidade Federal do Rio Grande do Sul, CP 15051, Porto Alegre 91501-970, RS, Brazil*
[6] *Universidade de São Paulo, Instituto de Astronomia, Geofísica e Ciências Atmosféricas, Rua do Matão 1226, CEP 05508-090, São Paulo, SP, Brazil*

**ABSTRACT**
We analyze the validity of optical diagnostic diagrams relying on emission-lines ratios and in the context of classifying Active Galactic Nuclei (AGNs) according to the cosmic metallicity evolution in the redshift range $0 \leq z \leq 11.2$. In this regard, we fit the results of chemical evolution models (CEMs) to the radial gradients of the N/O abundances ratio derived through direct estimates of electron temperatures ($T_e$-method) in a sample of four local spiral galaxies. This approach allows us to select representative CEMs and extrapolate the radial gradients to the nuclear regions of the galaxies in our sample, inferring in this way the central N/O and O/H abundances. The nuclear abundance predictions for theoretical galaxies from the selected CEMs, at distinct evolutionary stages, are used as input parameters in AGN photoionization models built with the Cloudy code. We found that standard BPT diagnostic diagrams are able to classify AGNs with oxygen abundances $12+\log({\rm O/H}) \gtrsim 8.0 \, [(Z/Z_\odot) \gtrsim 0.2]$ at redshift $z \lesssim 4$. On the other hand, the He ɪɪ$\lambda 4685$/H$\beta$ versus [N ɪɪ]$\lambda 6584$/H$\alpha$ diagram produces a reliable AGN classification independent of the evolutionary stage of these objects.

## 1 INTRODUCTION

Spectral classification of galaxies based on emission line intensities is fundamental for the understanding of the physical processes actuating in the Interstellar Medium (ISM) as well as for the knowledge of the ionizing source, of elemental abundances and of the metallicity of the gas phase of these objects.

Baldwin et al. (1981), in their pioneering study, showed that combinations of strong optical emission line intensities can be used to separate objects according to their main excitation mechanism: H ɪɪ regions or Star Forming galaxies (hereafter SFs), Planetary Nebulae, Active Galactic Nuclei (AGNs), and objects excited by shock-waves. These authors proposed the [O ɪɪɪ]$\lambda 5007$/H$\beta$ versus [N ɪɪ]$6584$/H$\alpha$ diagnostic diagram to distinguish SFs from other object classes. Hereafter, this diagram is named as [N ɪɪ]-diagram. Veilleux & Osterbrock (1987) revised the method of classification proposed by Baldwin et al. (1981), including the [S ɪɪ]($\lambda 6716+\lambda 6731$)/H$\alpha$ and [O ɪ]$\lambda 6300$/H$\alpha$ line ratios, commonly known as BPT diagrams. The separation lines between SFs and AGNs in BPT diagrams, called as maximum starburst lines, have been mainly established by using observational data of objects in the local universe ($z<1$, e.g. Kauffmann et al. 2003) and relied on photoionization model results (e.g. Kewley et al. 2001). However, optical spectroscopic data of SFs at high redshift ($z>1$) have shown an offset of the maximum starburst lines from the local star-forming ones in the [N ɪɪ]-diagram, in the sense that SFs at higher redshifts can show emission-line intensity ratios commonly presented by AGNs (e.g. Shapley et al. 2005, 2015; Liu et al. 2008; Hainline et al. 2009; Bian et al. 2010; Steidel et al. 2014; Masters et al. 2014; Sanders et al. 2016a; Strom et al. 2017; Kashino et al. 2017; Cameron et al. 2023; Simmonds et al. 2023).

The cause of this offset is under discussion and it has been attributed to higher ionization parameters, harder radiation fields, higher N/O ratios in high redshift SFs and/or due to a bias in sample selection (see Sanders et al. 2023a; Garg et al. 2022). For instance, Kewley et al. (2013) compared a large sample of galaxies in the redshift range $0.5<z<2.6$ with predictions of photoionization models in the [N ɪɪ]-diagram. These authors concluded that the offset is due to the ISM conditions are more extreme at high redshift than those of local galaxies (see also Sugahara et al. 2022). Otherwise, Sanders et al. (2016b), using observations of confirmed SFs (identified by X-Ray or infrared properties) at $z\sim 2.3$ taken from the MOSFIRE Deep Evolution Field survey (Kriek et al. 2015), found that the cause of the offset of high-$z$ SFs in the [N ɪɪ]-diagram is due to the elevated N/O abundance ratio at fixed O/H (see also Shapley et al. 2015; Masters et al. 2014; Pérez-Montero & Contini 2009). Moreover, a combination of metallicity and ionization parameter effects can be the cause of the maximum starburst line displacement (Cowie et al. 2016; Garg et al. 2022; Papovich et al. 2022).

⋆ E-mail: olidors@univap.br

Regarding AGNs, this object class can occupy the SF zone in the [N II]-diagram even in the local universe, yielding an incorrect classification of galaxy nuclei. Groves et al. (2006), who used photoionization models to examine the effects of metallicity ($Z$) variations on the narrow emission lines of AGNs, showed that AGNs with $(Z/Z_\odot) \lesssim 0.5$ can occupy similar regions than SFs with high ionization [log([O III]$\lambda$5007/H$\beta$) $\gtrsim 0.5$]. In the same direction, Feltre et al. (2016) compared SF and AGN photoionization model results with observational data of a large sample of objects ($z < 0.2$) taken from Sloan Digital Sky Survey Data Release 7 (SDSS-DR7; Abazajian et al. 2009) in order to analyze the discrimination by the use of line-ratio diagnostics. These authors found that for the low metallicity regime [$(Z/Z_\odot) \lesssim 0.4$], AGN and SF models predict similar values of [O III]$\lambda$5007/H$\beta$ and [N II]6584/H$\alpha$, on the side of the [N II]-diagram corresponding to SFs (see also Feltre et al. 2023). Thus, despite there is evidence that only SFs at high-$z$ are located above the maximum starburst line, AGNs in the local universe and with low metallicity can occupy the SF zone in BPT diagrams (see also Zhu et al. 2023). This fact produces a bias in the AGN selection and in statistic metallicity studies because active galaxies with low metallicity, misclassified as SFs, are excluded from any selection/analysis. In fact, Osorio-Clavijo et al. (2023) used optical data taken from the Calar Alto Legacy Integral Field Area survey (CALIFA, Sánchez et al. 2012) and data from the CHANDRA X-ray satellite (Weisskopf et al. 2002) to study the AGN population in the local universe. These authors showed that the AGN population in the CALIFA survey is 4 per cent (Lacerda et al. 2020) when only BPT diagrams are used as a selection criterion. However, this fraction could rise up to $\sim$ 10 per cent when accounting for both optical and X-ray spectroscopic analyses. Also, Bykov et al. (2023) showed that most part of AGNs hosted in low mass galaxies ($M < 10^{9.5}$ M$_\odot$) and selected by using X-ray luminosity are located in the SF zone in the [N II]-diagram. Moreover, Dors et al. (2020) selected 463 AGNs ($z < 0.4$) based on BPT diagrams using the SDSS-DR7 (Abazajian et al. 2009) database and classifications from NASA/IPAC Extragalactic Database (NED). These authors estimated the metallicity through several strong-line methods and found that AGNs with $(Z/Z_\odot) < 0.6$ are only $\sim$ 10 per cent of the sample. Certainly, if standard BPT diagrams exclude low metallicity AGNs the fraction above is underestimated.

Important insights on the applicability of BPT diagrams at high-$z$ are coming from sophisticated analysis relying on hydrodynamic simulations combined with photoionization models. For instance, Garg et al. (2022), by using the SIMBA code (Davé et al. 2019), simulated galaxies contained in a box with a side length of 100 $h^{-1}$ Mpc. The abundances resulting from SIMBA simulations are assumed as input in SF photoionization models. From this simulation, Garg et al. (2022) showed that SFs at $z = 2$ either with an increasing of the N/O abundance ratio or with a decreasing ionization parameter at fixed O/H can go through the maximum starburst line in the [N II]-diagram. Hirschmann et al. (2022) combined cosmological ILLUSTRISTNG[1] simulations (Nelson et al. 2018) with SF and AGN photoionization models, gas ionization by post-AGB stars as well as fast, radiative shocks in order to analyze the validity of diagnostic diagrams. These authors showed that standard BPT diagrams break down for metal-poor ($Z \lesssim 0.5 Z_\odot$) AGNs at $z \gtrsim 1$ (see also Hirschmann et al. 2019; Nakajima & Maiolino 2022), hence AGN galaxies move towards the SF zone.

As pointed out by Hirschmann et al. (2022), a source of the inaccuracy of current large-scale cosmological simulations (see also Buck et al. 2021), such as the ILLUSTRISTNG and SIMBA simulations, is that they do not resolve the ISM, i.e. these simulations neither resolve gas properties, such as density in individual ionized regions, nor track the formation, evolution and destruction of dust grains. This problem makes that observed AGN luminosity function, stellar populations and associated scaling relations are not always well reproduced by these models. To circumvent the possible limitation of cosmological hydrodynamic simulations, detailed fitting of chemical evolution models to physical parameter estimations (e.g. star formation rate, abundances, etc) of individual galaxies can produce more reliable constraints to the applicability of the BPT diagrams. For instance, Mollá & Díaz (2005) presented a generalization of the multiphase chemical evolution model (hereafter CEM) applied to a wide set of theoretical galaxies with different masses and evolutionary rates. Over decades, this classical approach of CEMs have been able to predict the radial elemental abundance distributions (with a spatial resolution of few kpc) of galaxies (e.g. Cavichia et al. 2023; Sharda et al. 2021; Mollá et al. 2019; Magrini et al. 2016; Kang et al. 2016; Kubryk et al. 2015; Cavichia et al. 2014; Marcon-Uchida et al. 2010; Schönrich & Binney 2009; Chiappini et al. 2001; Prantzos & Boissier 2000; Diaz & Tosi 1984; Talbot 1980; Alloin et al. 1979).

Another approach of most simulations built to analyze the validity of diagnostic diagrams along the cosmic time is the assumption of the O/H abundance as a metallicity tracer of the ISM (e.g. Garg et al. 2022). However, for AGN photoionization modeling this approach can introduce some uncertainty in the O/H abundance or metallicity values and, consequently, in the predicted emission line intensities used in the BPT diagrams. Radio observations have shown the existence of neutral/molecular gas reservoir in the central parts of galaxies containing AGNs (e.g. Dressel et al. 1982; Hutchings et al. 1987; Bertram et al. 2007; Ho et al. 2008; García-Burillo et al. 2014; Bradford et al. 2018; Ellison et al. 2019; Combes et al. 2019). Recently, do Nascimento et al. (2022), by using MaNGA spectroscopic data of 108 Seyfert nuclei, showed that this neutral/molecular gas reservoir seems to be the cause of the lower (0.16-0.30 dex) O/H abundance in AGNs than the expected value from central extrapolation of radial metallicity gradients. Likewise, elemental abundance estimates (e.g. O/H, N/H) can be somewhat uncertain due to several factors. For example, O/H abundances based on direct estimates of electron temperature ($T_e$-method) can be underestimated (up to $\sim$ 0.2 dex) due to electron temperature fluctuations, possibly present in SFs (e.g. Méndez-Delgado et al. 2023; Oliveira et al. 2008; Peimbert et al. 2007; Hägele et al. 2006; Krabbe & Copetti 2002) and in AGNs (Riffel et al. 2021). On the other hand, it is largely known that photoionization models trend to overestimate elemental abundances (by 0.1-0.4 dex) in relation to those via the $T_e$-method (e.g. Kennicutt et al. 2003; Dors & Copetti 2005; Kewley & Ellison 2008). Thus, chemical evolution model predictions and elemental abundance estimations of galaxies involving hydrogen content (e.g. O/H, N/H) can be somewhat uncertain. In this sense, the abundance ratio of heavy elements (e.g. N/O, C/O, Fe/O) trend to be more reliable metallicity indicators in comparison to O/H abundance, since they do not take into account the hydrogen content, as well as the effects of electron temperature fluctuations and the uncertainties in photoionization models trend to be minimized. In fact, Watanabe et al. (2023) compared the values of the abundance ratios (Fe/O, Ar/O, S/O) predicted by chemical evolution models with direct abundance estimates for galaxies in the local universe and in the range of $z \sim 1-10$. Important results arose from this comparison as, for instance, winds of rotating Wolf Rayet stars that end up as a direct collapse could explain the high N/O abundance [log(N/O) $> -0.5$] derived for the galaxy GN-z11 ($z \sim 10.6$, Cameron et al. 2023; Senchyna et al. 2023) rather

---
[1] https://www.tng-project.org/

than supernovae ISM enrichment (but see Maiolino et al. 2023 for a different result).

In particular, the N/O abundance ratio has been suggested as a good metallicity tracer and it is correlated to the galaxy mass (e.g. Hayden-Pawson et al. 2022; Florido et al. 2022; Kojima et al. 2017; Douglass & Vogeley 2017; Pérez-Montero et al. 2016; Masters et al. 2016; Andrews & Martini 2013; Pérez-Montero & Contini 2009) as well as it is a useful tool to study the interplay of galactic processes, for instance, star formation efficiency, the timescale of infall, and outflow (e.g. Johnson et al. 2023; Magrini et al. 2018). In addition, recent estimates of radial abundance gradients in local spiral galaxies via direct estimations of the electron temperatures ($T_e$-method[2]), for example, make available by the CHAOS project[3] (e.g. Rogers et al. 2022, 2021; Croxall et al. 2016, 2015; Berg et al. 2015) as well as estimates for very high redshift ($z > 5$) galaxies obtained by the James Webb Space Telescope (e.g. Curti et al. 2023b; Arellano-Córdova et al. 2022; Hsiao et al. 2023) provide reliable constraints to chemical evolution models and, consequently, to the validity of the BPT diagrams in the cosmological context.

Motivated by the large number of direct elemental abundance estimates in galaxies with a wide range of redshift and by the advantage of the use of N/O as a metallicity tracer, in this work, we used predictions of the chemical evolution models proposed by Mollá & Díaz (2005) to reproduce direct estimations of N/O gradients in four local spiral galaxies, whose data were taken from the literature. Extrapolations of the radial gradients to the central parts of the galaxies, predicted by the selected models by Mollá & Díaz (2005), were carried out in order to obtain the central N/O abundances, in a wide range of redshifts ($z = 0 - 11$). Furthermore, the resulting N/O and the corresponding O/H abundances, for different redshift bins, were assumed as input parameters in AGN photoionization models built with the CLOUDY code (Ferland et al. 2013). This methodology makes it possible to investigate the feasibility of optical diagnostic diagrams for the AGN classification in a wide range of redshifts. The structure of this paper is as follows. In Section 2 the methodology employed in the analysis, i.e. observational abundance estimates used as constraints, CEMs and photoionization models descriptions, is presented. In Sect. 3 the results and their discussion are presented. Conclusions are presented in Sect. 4. Throughout this paper we adopt the Planck Collaboration et al. (2021) cosmologic parameters: $H_0 = 67.4$ km s$^{-1}$ Mpc$^{-1}$ and $\Omega_m = 0.315$.

## 2 METHODOLOGY

We obtained optical emission line intensities predicted by photoionization models in order to analyze the reliability of the BPT diagrams for a wide redshift range. For that, we used predicted N/O radial abundances from CEMs proposed by Mollá & Díaz (2005) to reproduce the ones derived through the $T_e$-method in a small sample of local spiral galaxies. After that, the N/O gradient resulting for each theoretical galaxy with distinct evolutionary stages was extrapolated to its central part (galactocentric distance equal to zero) in order to obtain its nuclear abundance. The N/O nuclear abundances and their corresponding O/H abundances were used as input parameters in AGN photoionization models. Finally, the optical emission-line intensities ratios predicted by the photoionization models were employed in order to investigate the galaxy classification for the redshift range

---

[2] For a review of the $T_e$-method see Peimbert et al. (2017) and Pérez-Montero (2017).
[3] https://www.danielleaberg.com/chaos

$z = 0.0 - 11.0$. In the following sections, we present each employed procedure.

### 2.1 Direct radial abundance estimates

We compiled from the literature the N/O radial abundances of the disks of four local ($z < 0.1$) spiral galaxies derived through direct measurements of the electron temperatures, i.e. by using the $T_e$-method, known as the most reliable method (see e.g. Pilyugin 2003; Hägele et al. 2006, 2008, 2012; Toribio San Cipriano et al. 2017). Since the $T_e$-method requires the presence of faint (about 100 times fainter than H$\beta$) auroral lines (e.g. [O III]$\lambda$4363; Díaz et al. 2007), direct estimates of N/O radial gradients are only available for a few spiral galaxies and do not cover the entire optical disks. In fact, only for 20 over the 41 disk H II regions belonging to the grand design and nearby spiral galaxy M 101 (NGC 5457) observed by Kennicutt & Garnett (1996), it was possible to measure electron temperatures (see Kennicutt et al. 2003). Likewise, more recent spectroscopic data of spiral galaxies obtained by the CHAOS project (e.g. Berg et al. 2015), a project dedicated to measuring direct abundance in disk H II regions using the Large Binocular Telescope, have shown the difficulty to measure electron temperatures even in nearby galactic disks. Thus, to assume only abundance estimates via the $T_e$-method as a selection criterion penalizes any study, in the sense that few objects are able to be selected (e.g. Dors et al. 2020).

To estimate the radial N/O abundance distributions of spiral galaxies, it is usually assumed a linear regression:

$$\log(\mathrm{N/O}) = \log(\mathrm{N/O})_0 + grad\,\log(\mathrm{N/O}) \times R, \qquad (1)$$

where $R$ is the galactocentric distance (in units of kpc), $\log(\mathrm{N/O})_0$ is the extrapolated value of this parameter to the galactic centre ($R = 0$ kpc) and *grad* is the slope (in units of dex kpc$^{-1}$). To derive the N/O galactic gradient, H II regions distributed across the galaxy disk are necessary (e.g. Gusev et al. 2012). Therefore, we selected from the literature only galaxies for which N/O estimates via the $T_e$-method were obtained for more than 10 disk H II regions with galactocentric distances $R$ reaching up to the isophotal radius $R_{25}$ (see Pilyugin et al. 2004), defined as the galactocentric distance where the surface brightness is 25 mag arcsec$^{-2}$ in the *B*-band. The selected objects, the number of disk H II regions and the authors who calculated the N/O estimates are: NGC 628 (45, Berg et al. 2015), NGC 2403 (28, Rogers et al. 2021), M 101 (20, Kennicutt et al. 2003), and M 33 (60, Rogers et al. 2022).

The following methodology was applied by the authors (see Table 1) to estimate the direct abundances:

• Two electron temperatures [$T_e$(low) and $T_e$(high)] are derived through the [N II]($\lambda$6548 + $\lambda$6584)/$\lambda$5755 and [O III]($\lambda$4959 + $\lambda$5007)/$\lambda$4363 line intensities ratios for the zones were the O$^+$ - N$^+$ and the O$^{+2}$ ions are originated, respectively (Peimbert 1967). For the cases when it was not possible to apply the previously mentioned procedure to estimate $T_e$(low), it was estimated through theoretical/empirical relations with the $T_e$(high) (see e.g. Hägele et al. 2008; Rogers et al. 2022).

• The total oxygen and nitrogen abundances are estimated assuming (O/H)=(O$^+$/H$^+$)+(O$^{2+}$/H$^+$) and (N/O)=(N$^+$/O$^+$) (Peimbert & Costero 1969).

It is worth mentioning that as the N/O estimates for the objects in our sample were taken from different authors, for which distinct observational techniques, reddening law and atomic parameters were assumed, the estimates are obtained adopting not a homogeneous methodology. This could produce artificial scattering or biases in the derived abundances and, consequently, in the CEM interpretations. Effects of the use of distinct methodologies and heterogeneous samples on nebular abundance analysis are discussed, for instance, by Juan de Dios & Rodríguez (2017) and Dors et al. (2023), which could introduce abundance variations in the order of ∼ 0.1 dex, which is similar to the uncertainties introduced by the errors in the optical emission-line measurements (see e.g. Hägele et al. 2008).

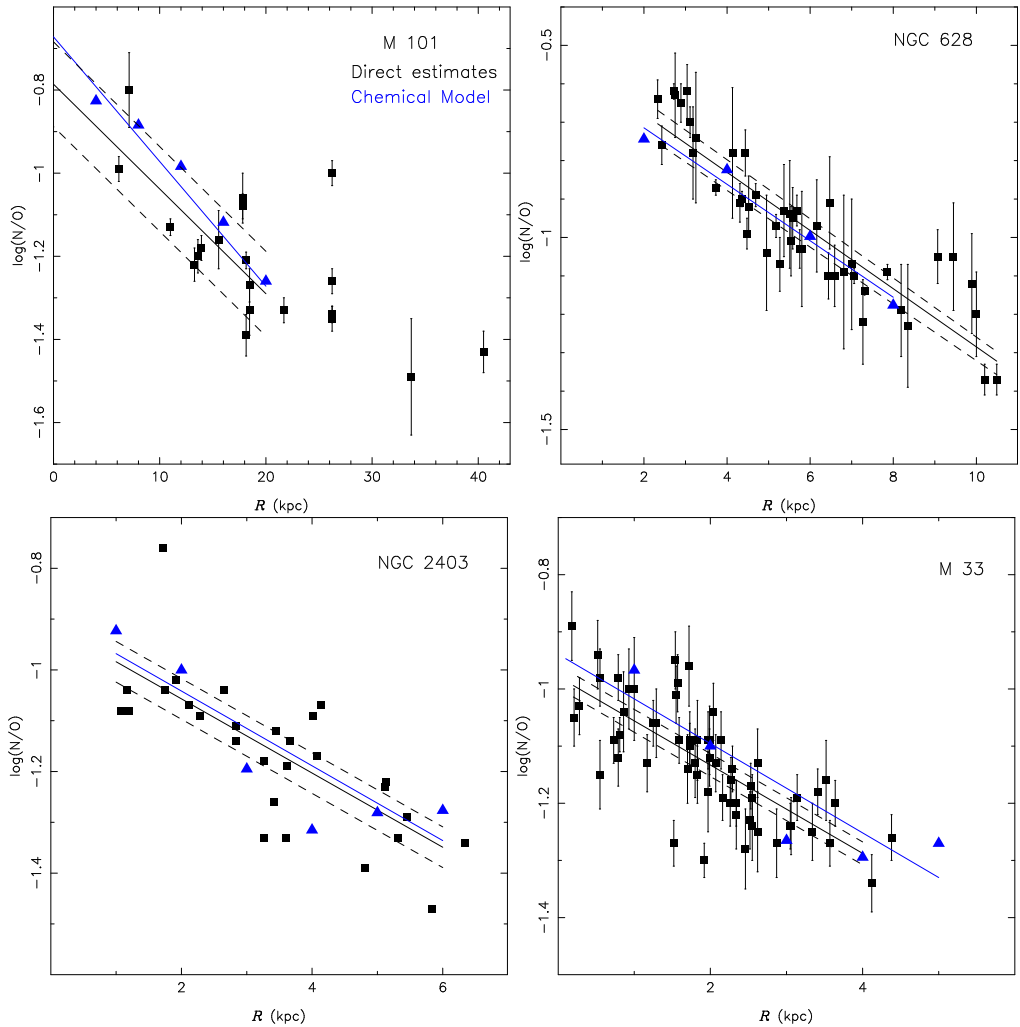

**Figure 1.** Comparison between N/O abundance ratio estimates based on the $T_{\rm e}$-method (black points) and predictions from chemical evolution models by blue points 2005MNRAS.358..521M for the spiral galaxies indicated on each panel. Solid black lines represent linear regressions to the estimates via the $T_{\rm e}$-method while dashed black lines the uncertainties of these fittings. Blue lines represent linear regressions to the chemical evolution model results. Table 1 lists the main features of the considered galaxies and the references for direct abundance estimates.

In Figure 1, the logarithm of the direct N/O abundances taken from the literature as a function of the galactocentric distances are plotted together with their linear regressions obtained following the Equation 1 for each studied galaxy. The fitting uncertainties are also shown.

In Table 1, the identification of the spiral galaxies considered in our analysis, the logarithm of their masses (in units of the solar mass), the number of disk H II regions with direct abundance estimates, the galactocentric distance $R$ range in which the N/O values were derived, and their literature references are presented.

Zaritsky et al. (1994) showed the existence of correlations between the O/H gradients and galactic macroscopic quantities (e.g. mass, absolute magnitude, Hubble type), being these correlations dependent on the normalization to the galactocentric distance (see also Vila-Costas & Edmunds 1992; Pilyugin et al. 2004, 2019; Ho et al. 2015). In particular, less massive galaxies trend to have steeper gradients (in units of dex kpc$^{-1}$) in comparison to those in more massive objects (e.g. Ho et al. 2015). Despite this result is still under debate in the literature (see e.g. Belfiore et al. 2017; Poetrodjojo et al. 2018), any study aiming to obtain representative abundance gradients in spiral galaxies must consider a wide galactic mass interval (e.g. Hirschmann et al. 2022). The mass interval of our sample is $9.6 \lesssim \log(M/M_\odot) \lesssim 11$, covering practically all the mass range considered by Maiolino et al. (2008), who derived the mass-metallicity relation for galaxies in a wide redshift range. Even though abundance values and metallicity gradients for galaxies with mass values lower than $\log(M/M_\odot) \lesssim 9.0$ are available in the literature (e.g. Ho et al. 2015; Bresolin 2019; Curti et al. 2023a), N/O gradients via the $T_{\rm e}$-method in these objects are barely found.

In addition, galaxies located in dense environments (e.g. Dors & Copetti 2006; Mouhcine et al. 2007; Robertson et al. 2012; Lara-López et al. 2022) as well as interacting galaxies (e.g. Krabbe et al. 2008, 2011; Kewley et al. 2010; Rupke et al. 2010; Sánchez et al. 2014; Rosa et al. 2014; Torres-Flores et al. 2014; Zinchenko et al. 2015) trend to have different metallicities and, consequently, abun-dance gradients in comparison to the isolated ones. Moreover, the shape of oxygen abundance profiles, in some cases, can not be represented by a single function (e.g. Martin & Roy 1995; Roy & Walsh 1997; Bresolin et al. 2009, 2012; Marino et al. 2012; Pilyugin et al. 2017; Sánchez-Menguiano et al. 2018), as the one proposed in the Eq. 1. Our sample of spiral galaxies do not present clear signals of interactions with near galaxies, as reported in the works from which the direct estimates were taken. Likewise, as we can see in Fig. 1, for 3/4 objects, the N/O abundance distributions are well represented by a linear fitting, with slopes steeper than $-0.02$ dex kpc$^{-1}$, not indicating any effect of environment/interaction on the abundances. This feature is important because the CEMs proposed by Mollá & Díaz (2005) consider only the cosmic evolution of isolated galaxies.

For 3/4 galaxies in our sample, clear O/H and N/H single linear regressions with negative slopes were derived in the works from which the estimates were obtained (not shown), and thus, the N/O regressions shown in Fig. 1 indicate that the nitrogen has mainly a secondary source of production, i.e. $A(N) \sim A(O)^2$ (e.g. Hamann & Ferland 1993), being $A$ the abundance. It can be also seen in Fig. 1, that for the H II regions belonging to M 101 and located at galactocentric distances larger than $\sim 20$ kpc, a flattened N/O distribution is derived. This result could be due to these H II regions having low metallicities [$12 + \log(O/H) \lesssim 8.0$], and for this regime the nitrogen has mainly a primary origin (Kennicutt et al. 2003). For the other three galaxies, we are sampling the inner part of the disk and they show similar behavior to M 101 for $R \lesssim 20$ kpc. For consistency, in M 101, we only consider N/O estimates for the 13 regions with $R \lesssim 20$ kpc, i.e. only those estimates showing a secondary nitrogen production.

### 2.2 Chemical evolution models

Mollá & Díaz (2005) presented multiphase chemical evolution models applied to a wide set of theoretical galaxies with different masses ($9 \lesssim \log(M/M_\odot) \lesssim 13$), evolutionary rates (age ranging from $\sim 0$ to $\sim 13$ Gyr) and distinct efficiencies of stellar formation. In these models, the enriched material proceeds from the restitution by dying stars, considering their nucleosynthesis predicted by the stellar products of massive stars by Woosley & Weaver (1995), low and intermediate-mass stars by Gavilán et al. (2005), and an initial mass function (IMF) by Ferrini et al. (1990) that is similar to the one proposed by Scalo (1986). These chemical evolution models predict elemental abundances (e.g. O, N, Fe, etc) along the hypothetical galaxy disks assuming that this is formed by radial regions, being the smallest $R$ value of 1-4 kpc, depending on the galaxy mass.

To fit the CEM results to the direct radial abundance for our sample of spiral galaxies the following methodology was adopted.

• Galaxy Mass: The maximum galactic radius ($R_{gal}$) is correlated to the galaxy mass ($M_{gal}$) by the expression (Lequeux 1983):

$$M_{gal} = 2.32 \times 10^5 \, R_{gal} \, V_{max}^2, \quad (2)$$

being $V_{max}$ the maximum rotation velocity. Initially, we selected a set of models ($S1$) whose predicted N/O abundances cover the range [$\Delta(N/O)$] of direct estimates along the radius ($R$) of each galaxy of the sample, not taking into account a fitting to the abundance ratio gradient (see also Magrini et al. 2016). Basically, we consider

$$(S1) = \Delta(N/O)_{CEM} \approx \Delta(N/O)_{dir.}, \quad (3)$$

being $\Delta(N/O)_{CEM}$ and $\Delta(N/O)_{dir.}$ the range of abundance ratios predicted by the CEMs and derived by the $T_e$-method, respectively. Thus, this fitting procedure allows us to find the best set of CEMs

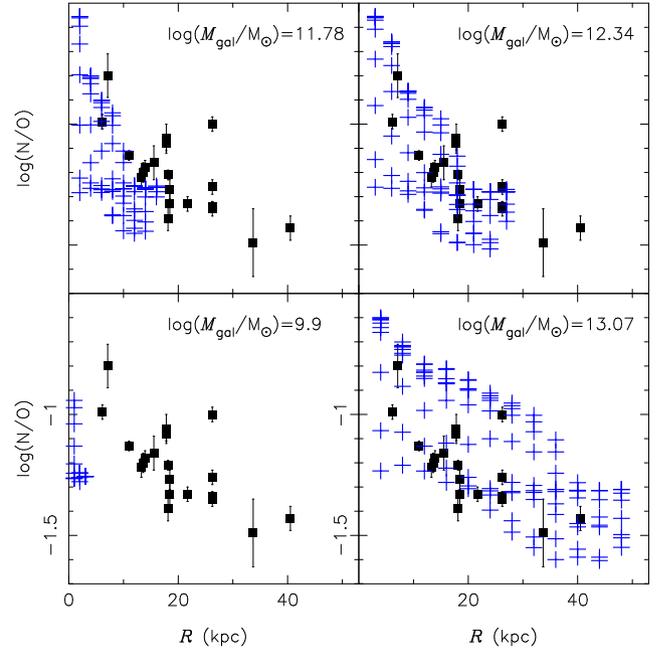

**Figure 2.** Logarithm of the N/O abundance ratio versus the galactocentric distance $R$ (in kpc) for the M 101 galaxy. Black points represent direct abundance estimates compiled from the literature (see Table 1). In each panel, blue crosses represent predictions from CEM by Mollá & Díaz (2005) for distinct galaxy mass ($M_{gal}$), as indicated, for all star formation efficiencies (not discriminated) and assuming an age of 13.2 Gyr.

($S1$) with distinct masses that describe the direct radial abundances. For each galaxy of our sample, $S1$ contains one (unique solution) or at maximum two CEMs with distinct mass. For instance, in Fig. 2, the direct abundance estimates compiled from the literature (see Table 1) for M 101 galaxy is compared with CEM predictions by Mollá & Díaz (2005) assuming four $M_{gal}$ values (which cover the galaxy mass interval for the CEMs) and taking into account all star formation efficiencies (not discriminated). This grand design spiral galaxy has served as the prototype system for studies on gas-phase abundances in extragalactic objects (Kennicutt & Garnett 1996). We can see in Fig. 2 that the CEM with $\log(M_{gal}/M_\odot) = 13.07$ describes clearly the direct estimates.

• Collapse time-scales ($\tau$): the $\tau$ value for each galaxy depends on its total mass through the expression given by Gallagher et al. (1984):

$$\tau \propto M_{gal}^{-1/2} \, T, \quad (4)$$

being $T$ the age, i.e. galaxies with lower masses have higher $\tau$ than more massive ones. The age of a given galaxy is difficult to estimate because it requires the chronochemodynamics of stars to reconstruct a timeline of galaxy events (for a review see Soderblom 2010). As there are no age estimations for the galaxies in our sample, we assume only chemical models with 13.2 Gyr. Recent James Webb Space Telescope (JWST) observations have revealed the presence of galactic disks at $3 \lesssim z \lesssim 12$, corresponding to $\sim 2$ and $\sim 0.3$ Gyr, respectively, after the Big Bang (e.g. Ferreira et al. 2022a,b; Naidu et al. 2022; Robertson et al. 2023). Thus, it seems to be a good approach to assume for our sample ($z \sim 0.0$) ages of 13.2 Gyr.

• Star formation efficiencies ($\epsilon$): the CEMs by Mollá & Díaz (2005) assume star formation that takes place in two steps: first, the formation of molecular clouds with efficiency $\epsilon_\mu$; then the cloud–cloud collision with efficiency $\epsilon_H$. Both $\epsilon_\mu$ and $\epsilon_H$ are re-lated to the star formation rate (SFR, Mollá & Díaz 2005). The star formation efficiency is defined by:

$$\epsilon = \frac{\ln \epsilon_\mu}{\ln \epsilon_H}, \quad (5)$$

with $\epsilon$ constant along the galactic disk. Mollá & Díaz (2005) computed 10 models for each theoretical galaxy, allowing $\epsilon_\mu$ ranging from 0 to 1 and $\epsilon_H$ fixed according to the ratio ($\frac{\ln \epsilon_\mu}{\ln \epsilon_H}$) ~ 0.4. We follow the designation presented in Table 2 of Mollá & Díaz (2005), where CEMs with different $\epsilon_\mu$ values are defined by the ¨$N$¨ term.

Mollá & Díaz (2005) built 440 different models that represent all possible combinations of the parameters described above.

Taking into account the set $S1$, we establish a new set of CEMs ($S2$), now, comparing the predictions and direct N/O gradient parameters for each galaxy of the sample. To fit the CEM results to the direct radial N/O estimates, for each galaxy of our sample, we assume that a CEM is representative if it yields the minimum value of $\chi$:

$$\chi^2(S2) = f(M_{\rm gal}, N, \tau = 13.2\,{\rm Gyr}) = \chi_c^2 - \chi_g^2, \quad (6)$$

being $\chi_c = \log({\rm N/O})_0^{\rm ob.} - \log({\rm N/O})_0^{\rm mod.}$, $\chi_g = grad\,\log({\rm N/O})_{\rm ob.} - grad\,\log({\rm N/O})_{\rm mod.}$, where the ¨ob.¨ and ¨mod.¨ indices indicate the observational and chemical model linear regressions parameters, respectively. The selected CEM results are plotted in Fig. 1 together with their corresponding linear regressions. A unique CEM is found to be representative of each galaxy.

In Fig. 3, direct abundance estimations [log(N/O)] as a function of galactocentric distances ($R$) are compared with CEM predictions by Mollá & Díaz (2005) for the M 101 galaxy. The CEM predictions are for distinct $N$ values and fixed $M_{\rm gal}$ and $\tau$ values, as indicated. From this comparison, we can see that the theoretical radial N/O distributions produce a bad fitting to the direct estimates for large $R$ values (see also Magrini et al. 2016; Mollá et al. 2019). The behaviour shown in Fig. 3 for M 101 is also derived for CEM predictions assuming other $M_{\rm gal}$ values (not shown). The discrepancy derived for large $R$ values, probably, is due to the simplistic approach that outer regions evolve at the same rate than inner galactic disk regions (e.g. Magrini et al. 2016), the effect of galactic interactions (more pronounced in outskirt regions, e.g. Toomre & Toomre 1972) and/or wrong SFR suppositions along the hypothetical disk (see Belfiore et al. 2017 and references therein). Recently, Garcia et al. (2023), by using gas-phase metallicity profiles of galaxies predicted by Illustris and IllustrisTNG simulations, showed that breaks in metallicity radial profiles are derived for galaxies at $z$ =0-3 and with a wide mass range, being these due to radial gas mixing. The existence of breaks in the radial N/O linear distributions pointed out the need to assume only linear regressions in the inner parts of the galactic disks in order to infer nuclear abundances, as carried out in the present work.

The comparisons shown in Figs. 2 and 3 clearly indicate the need to impose observational constraints on CEMs. Otherwise, parameters (e.g. masses, star formation efficiencies) no representative to those of real galaxies can be assumed in the CEMs and, consequently, wrong interpretations from these could be obtained.

We need to infer the central intercept N/O abundance [(N/O)$_0$] from the extrapolation of the linear regressions fitted to the theoretical values predicted by the CEMs and along the cosmic time. Due to the possible wrong theoretical predictions for large $R$ values, gradient predictions for the entire disks would produce incorrect (N/O)$_0$ values. To explore this fact, we study the behaviour of the logarithm of the (N/O)$_0$ obtained by fitting a linear regression (Eq. 1) to the CEM results (considering the models with $N$=8 as was selected for M 101 from Fig. 3) in the 0.0-11.2 redshift range (see Fig. 4). We fitted these linear regressions considering different galactocentric distances: $R$ < 20 kpc and the entire disk. We also considered, in Fig. 4, the (N/O) values for $R$ = 4 kpc as approximations for the (N/O)$_0$ values since this is the smallest galactocentric distance with abundance predictions by these CEMs and they are expected to have abundances near to those in nuclear regions. It can be noted in this figure that for $z \lesssim 6$ the log(N/O)$_0$ values estimated using the entire disks are ~ 0.3 dex lower than those estimated using $R$ < 20 kpc. Likewise, predictions assuming the log(N/O) at $R$ = 4 kpc as the log(N/O)$_0$ values are, as expected, somewhat lower (~ 0.15 dex) than those estimated via $R$ < 20 kpc. For $z \gtrsim 6$ all the suppositions produce similar (N/O)$_0$ values due to the flattened of gradients for high redshifts (see below). These results strengthen the assumption of only considering gradient estimates in disk parts where the nitrogen has mainly a secondary origin. Therefore, following this analysis performed for M 101, for the other galaxies in our sample (shown in Fig. 1), we only consider CEM predictions for the inner parts of their galactic disks, i.e. galactocentric distances similar to those for which direct estimates are available. In Table 1 the range of galactocentric distances considered in the CEM fittings are listed.

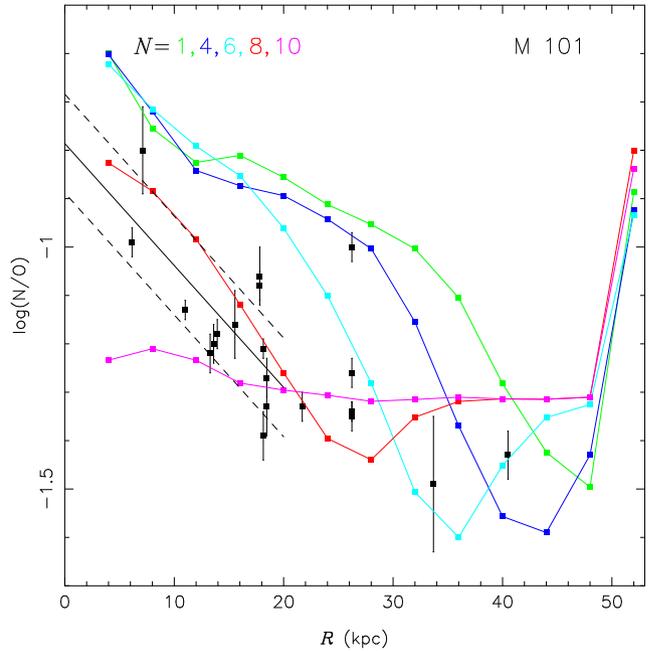

**Figure 3.** Logarithm of the N/O abundance ratio versus the galactocentric distance $R$ (in kpc) for the M 101 galaxy. Black points represent direct abundance estimates compiled from the literature (see Table 1) while solid and dashed black lines represent their fitting and its uncertainties, respectively. Color lines connect CEM results (colored points) for distinct star formation efficiency pairs $\epsilon_\mu$, $\epsilon_H$ according to their $N$ designation in Table 2 by Mollá & Díaz (2005) as indicated. The galactic mass [log($M_{\rm gal}/M_\odot$)] and collapse time-scale ($\tau$) are 13.07 and 13.2 Gyr, respectively.

### 2.3 Photoionization models

We considered the version 22.00 of the Cloudy code (Ferland et al. 2013, 2017) in order to build up photoionization model grids representing narrow line regions (NLRs) of AGNs in a wide redshift range ($z = 0.0 - 11.2$). In this regard, the (N/O)$_0$ values obtained from the linear fitting to the CEM radial gradient predictions, for dis-

**Table 1.** Properties of our sample of spiral galaxies and chemical evolution models. Columns: (1) Galaxy identification. (2) Galaxy mass. (3) Number of disk H ii regions with direct estimates of N/O abundance. (4) Galactocentric distance range with direct N/O estimates. (5) and (6) Slope and central intersect values of the N/O regression. (7) Redshift. (8) Scale. (9) Literature reference from which the observational data were taken. (10)-(11) model designations according to Table 2 by Mollá & Díaz (2005) (12) Galaxy mass derived through the corresponding CEM.

| | | | | | | | | | Chem. Mod. | | |
|---|---|---|---|---|---|---|---|---|---|---|---|
| (1) | (2) | (3) | (4) | (5) | (6) | (7) | (8) | (9) | (10) | (11) | (12) |
| Object | $\log(M_*/M_\odot)$ | Number | Radius (kpc) | $grad$(N/O) | $\log$(N/O)$_0$ | $z$ | Scale (kpc arsec$^{-1}$) | Ref. | N | $N$ | $\log(M_*/M_\odot)$ |
| M 101 | 10.7 | 13 | 6.00-20.00 | $-0.025 \pm 0.007$ | $-0.78 \pm 0.11$ | 0.00080 | 0.017 | 1 | 44 | 8 | 13.07 |
| NGC 628 | 10.0 | 45 | 2.00-11.00 | $-0.075 \pm 0.005$ | $-0.52 \pm 0.03$ | 0.00219 | 0.047 | 2 | 26 | 7 | 12.13 |
| NGC 2403 | 11.0 | 28 | 1.00-6.40 | $-0.073 \pm 0.012$ | $-0.91 \pm 0.04$ | 0.00044 | 0.009 | 3 | 14 | 7 | 11.24 |
| M 33 | 9.68 | 60 | 0.10-5.00 | $-0.080 \pm 0.023$ | $-0.93 \pm 0.07$ | $-0.00059$ | 0.004 | 4 | 10 | 7 | 11.07 |

References: (1) Kennicutt et al. (2003), (2) Berg et al. (2015), (3) Rogers et al. (2021), (4) Rogers et al. (2022).

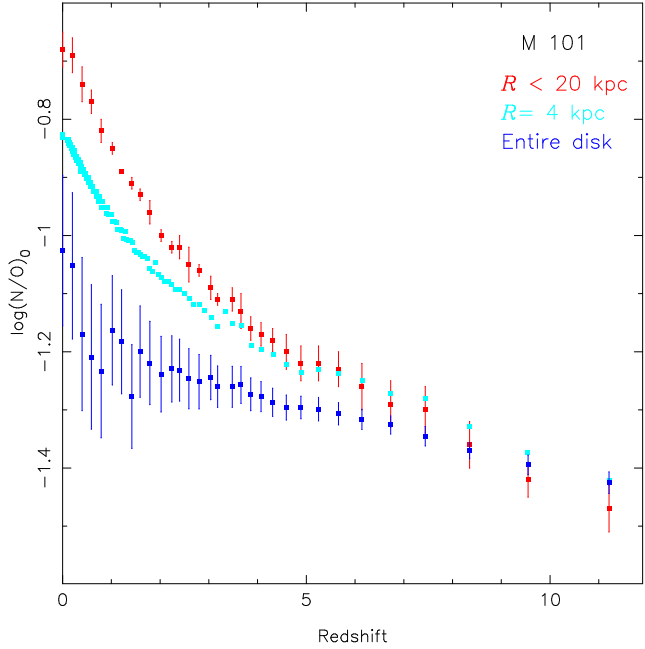

**Figure 4.** Logarithm of (N/O)$_0$ (extrapolated abundance to the galactic center) at different evolution times considering predictions from the CEMs by Mollá & Díaz (2005) for M 101 and assuming $N$=8 as selected through Fig. 3. Points represent (N/O)$_0$ assuming radial gradients estimated by fitting the Eq. 1 to the CEM predictions for the galactocentric distance $R < 20$ kpc, entire disk and assuming (N/O)$_0$ equal to (N/O) for $R = 4$ kpc (see text), as indicated. Error bars represent the uncertainty in (N/O)$_0$.

tinct redshifts and for each galaxy of our sample (see Sect. 2.1), and the correspondent O/H values (see below), were assumed as input parameters of the photoionization models.

These models were built following the methodology presented by Dors et al. (2019) and the reader is referred to this paper for a complete description. In what follows the input parameters are briefly described.

• Spectral Energy Distribution (SED): the SED was assumed to be the one defined by the Table AGN command in the Cloudy code and it is similar to that deduced by Mathews & Ferland (1987). We compared emission line intensities ratios, considered in BPT diagrams, predicted by our photoionization models with those assuming SEDs derived from observations of the AGNs Mrk 509 and NGC 5548 by Kaastra et al. (2011) and Mehdipour et al. (2015), respectively[4]. We found that photoionization models assuming these distinct SEDs produce line intensities ratios that differ by a factor lower than ∼ 0.1 dex (see also Feltre et al. 2016), similar to the uncertainties in observational line ratio intensities (e.g. Hägele et al. 2008).

• Electron density ($N_e$): in all photoionization models $N_e$ was assumed to be constant along the nebular radius and equal to 500 cm$^{-3}$, about the mean value found by Dors et al. (2014) for a large number of Seyfert 2 nuclei and derived through the [S ii]$\lambda$6716/6731 line ratio (see also Vaona et al. 2012; Zhang et al. 2013; Flury & Moran 2020; Agostino et al. 2021; XueGuang 2023). Armah et al. (2023) investigated the effect of the presence of a radial density profile (e.g. Congiu et al. 2017; Kakkad et al. 2018; Revalski et al. 2018; Freitas et al. 2018; Mingozzi et al. 2019; Revalski et al. 2021) and high density values ($N_e > 10^4$ cm$^{-3}$, see Vaona et al. 2012; Congiu et al. 2017; Cerqueira-Campos et al. 2021) in NRLs on photoionization model predictions. These authors found that density variations and high $N_e$ values yield changes in the emission-line intensities ratios used in BPT diagrams lower than 0.1 dex.

• Ionization parameter ($U$): we assume the logarithm of $U$ ranging from $-3.5$ to $-1.0$ dex, with a step of 0.5 dex. A similar range of $U$ was derived by Carvalho et al. (2020) from a comparison between photoionization model results and spectroscopic data of ∼ 430 Seyfert 2 (see also Pérez-Montero et al. 2019; Pérez-Díaz et al. 2022).

• Abundance set: the abundance of all metals ($Z$) was linearly scaled with the solar abundance[5] by a factor $f$, i.e. $Z = f \times Z_\odot$, with the exception of the nitrogen abundance. For each photoionization model representing a galaxy (see Sect. 2.1) at a given $z$, $Z$ is obtained from the (N/O)$_0$ value derived through the CEM using the N/O radial gradient.

Recently, Johnson et al. (2023), who proposed a multizone galactic chemical evolution model for Milky Way-like galaxies, showed that the (N/O)-(O/H) relation is dependent on the galactocentric distance, being this dependence mainly due to the pAGB stars evolution that enrich the ISM with N but with negligible amounts of O, increasing N/O. Basically, as spiral galaxies begin to form their inner regions before the outer ones in a classical inside-out scheme (e.g. Samland et al. 1997; Portinari & Chiosi 1999; Boissier & Prantzos 2000), the ISM at distinct $R$ values are enriched by stars with distinct evolutionary stages, resulting in a dependence of the (N/O)-(O/H) relation with $R$. In order to study this behavior using the assumed CEMs, we compare in Fig. 5 the theoretical (N/O)-(O/H) abundance ratios estimated for four distinct $R$ at distinct evolutionary stages ($z$ =0.0-11.2), and assuming the predictions for the M 101 CEMs. For simplicity, the CEM redshifts are not indicated. We can clearly see a discrepancy between the (N/O)-(O/H) relations, being that highest N/O values are only reached for the innermost ($R$ = 4 kpc) disk regions.

---
[4] These SEDs are implemented into Cloudy code.
[5] See the Hazy 1 manual of the Cloudy code (available in www.nublado.org) for abundance solar reference.

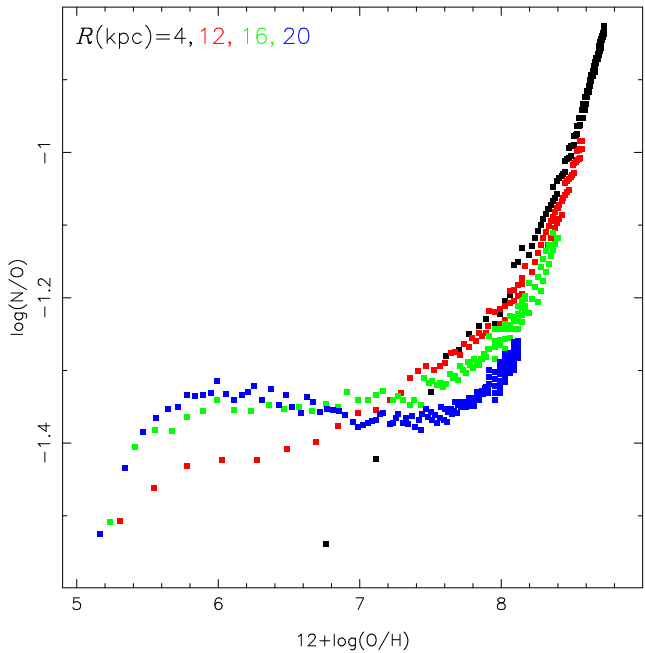

**Figure 5.** Logarithm of N/O abundance ratio versus 12+log(O/H). Points represent abundance predictions from the CEMs by Mollá & Díaz (2005) for distinct galactocentric distances, as indicated. CEMs were selected assuming the M 101 parameters (see Table 1 and for $z$=0-11.2 (not discriminated).

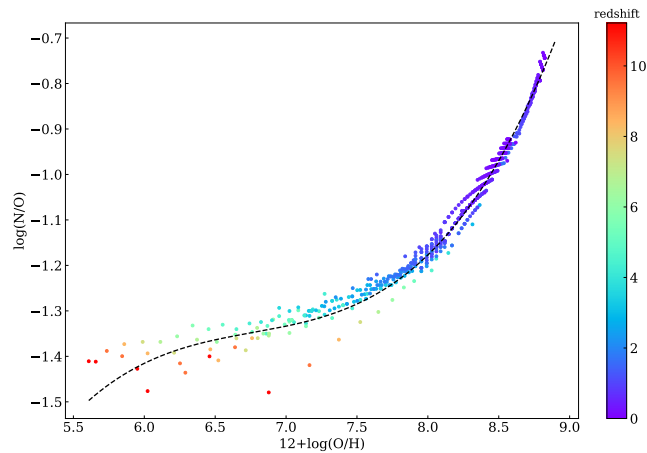

**Figure 6.** Logarithm of N/O abundance ratio versus 12+log(O/H). Points represent predictions of CEM abundances by Mollá & Díaz (2005) for the innermost galactocentric distances. The CEMs abundance values are listed in Tables A1-A4. The color bar indicates predictions for distinct redshifts, as indicated. The curve represents the fitting to the points given by Eq. 7.

Since we need to estimate the O/H abundance for the nuclear regions, we obtained the (N/O)-(O/H) relation for the inner galactocentric regions in each CEM [indicated in column (4) of Table 1] and assuming all redshift range ($z$ =0.0-11.2). In Fig. 6 this relation is shown discriminating their $z$, where a fitting to the points is given

by:

$$\log(N/O) = ax^3 + bx^2 + cx + d, \quad (7)$$

where x=12+log(O/H), a = 0.053839 ± 0.001943, b = −1.09321 ± 0.04405, c = 7.45593 ± 0.3306, and d = −18.4246 ± 0.8211.

For each model, the $f$ factor is defined as:

$$f = 10^{x_0 - 8.69}, \quad (8)$$

being 8.69 dex the solar oxygen abundance (Allende Prieto et al. 2001).

We use the $(N/O)_0$ abundance time evolution prescription by Mollá & Díaz (2005) for the chosen CEMs, listed in Tables A1-A4, and used as inputs in our photoionization models. To define the metallicity ($Z$) through the oxygen abundance $x_0$=12+log(O/H)$_0$, for each model representing a galaxy at a given $z$, the derived $(N/O)_0$ was used in Eq. 7 and, thus, its corresponding O/H abundance was obtained. We assume that Eq. 7 is still valid for $R$ = 0 kpc, and not only for the innermost galactocentric distances for which the CEMs were built. A similar, but inverse, procedure was adopted by Garg et al. (2022), where the O/H abundances were retrieved from the SIMBA simulations and N/O from observational estimates obtained by Pilyugin et al. (2012).

The helium abundance in relation to the hydrogen (He/H) was defined, in each model, by the expression

$$\begin{aligned}w =& (0.1215 \pm 0.0422) \times x^2 - (1.8183 \pm 0.6977) \times x \\ &+ (17.6732 \pm 2.8798),\end{aligned} \quad (9)$$

where w=12+log(He/H) and x=12+log(O/H). This expression was derived by Dors et al. (2022) through the $T_e$-method and by using a large number of AGN and H II region estimates (see also Dopita et al. 2006).

Recently, Zhu et al. (2023) compared dust-free and dusty AGN photoionization model results with observational AGN line intensities ratios in standard BPT diagrams (see also Feltre et al. 2016; Peluso et al. 2023). These authors found some significant discrepancy among the dust-free and dusty models only for the high metallicity regime, i.e. 12 + log(O/H) ≳ 8.7 [$(Z/Z_\odot)$ ≳ 1.0], with discrepancies for the line intensities ratios reaching up to 0.5 dex for 12 + log(O/H) = 9.2 [$(Z/Z_\odot)$ ∼ 3.2]. As the abundance of dust in the gas phase of gaseous nebulae is poorly known due to the difficulty of estimating this parameter (e.g. Sofia et al. 1994; Garnett et al. 1995; Peimbert & Peimbert 2010; Brinchmann et al. 2013; Martín-Doménech et al. 2016) and, since our models do not extend to the very high $Z$ regime, we followed Nagao et al. (2003, 2006b), who found a better match between dust-free AGN models to the observed emission line ratios in comparison to dusty models, and we assumed that all AGN photoionization models are dust-free.

## 3 RESULTS AND DISCUSSION

### 3.1 Galaxy mass and SF efficiency

In Fig. 1 the direct ($T_e$-method) N/O abundance estimations as a function of the galactocentric distances for the objects in our sample together with the results of the CEMs that better fit them are shown. In all the cases, it can be seen a good agreement between the linear fittings to the observational data and the CEM results, except for M 33 where the CEM fitting is offset by ∼ 0.05 dex from the one directly derived even though both fittings show the same slope. This difference is in the order of the uncertainties in abundance estimates via the $T_e$-method (∼ 0.1 dex, Kennicutt et al. 2003; Hägele et al.

2008) and lower than those via strong-line methods (~ 0.2 dex, e.g. Denicoló et al. 2002). Therefore, it has a small influence on the results obtained in the present work.

In Table 1 the CEM fitting parameters are listed. Masses of the theoretical galaxies are higher than those derived from observations by a factor up to ~ 2.4 dex. Discrepancies between observational and theoretical $M_{gal}$ estimations shown in Table 1 could be mainly due to only the inner parts of the galactic disks (the ones with direct N/O estimates) are considered in the CEM fittings, being any theoretical mass derivation through the CEMs highly uncertain.

We are interested in the nuclear N/O abundance, i.e. $(N/O)_0$, in order to provide it (and the corresponding O/H) as an initial parameter for our photoionization models and, thus, to investigate which abundances (and redshift) make AGNs trend to go through the maximum-starburst line in BPT diagrams. The $(N/O)_0$ values are obtained from radial gradients and they are dependent on the galaxy mass. Mollá & Díaz (2005) showed that O/H radial gradients are steeper for the models with lower maximum rotation velocities ($V_{max}$) or lower galaxy masses ($M_{gal}$, see also Belfiore et al. 2019). Consequently, for disk regions where nitrogen has mainly a secondary production, N/O gradients are also stepper when $M_{gal}$ decreases. Therefore, it is important that the range of $(N/O)_0$ derived for our sample be representative of spiral galaxies. In order to verify that, in Fig. 7 the distribution of $(N/O)_0$ values derived through the $P$-method (a method that yields abundances in agreement with those estimated via the $T_e$-method; Pilyugin 2001) for 45 nearby spiral galaxies by Pilyugin et al. (2004) are compared with the $(N/O)_0$ range predicted by CEMs for our sample of nearby galaxies ($z \sim 0$). The $\log(N/O)_0$ for our four spiral galaxies ranges from ~ −1.0 to ~ −0.5 dex, being ~ 70 per cent of the spiral galaxies studied by Pilyugin et al. (2004) have $\log(N/O)_0$ within this range. The extreme $(N/O)_0$ values estimated by Pilyugin et al. (2004) are not encompassed by our results. More N/O radial estimates via the $T_e$-method for spiral galaxies with a wide mass range could be necessary to extend our results.

Concerning the star formation efficiency $\epsilon$, we found that CEMs with $N$ equal to 7 and 8 (see Table 2 by Mollá & Díaz 2005) reproduce the radial N/O gradients of our sample. Magrini et al. (2016) also carried out fittings of the CEMs by Mollá & Díaz (2005) to O/H radial gradients of the spiral galaxies M 31, M 33 and NGC 300. These authors found that CEMs assuming the highest $\epsilon$ value (i.e. $N = 10$) are satisfactory to represent the gradients of their galaxies. These three $\epsilon$ values (i.e. $N = 7, 8$ and $10$) represent star formation with low efficiency in comparison to the $\epsilon$ representing the Milky Way (i.e. $N = 4$), according to Mollá & Díaz (2005). The source of this discrepancy is, probably, due to Mollá & Díaz (2005) considering a larger number of constraints in their models (e.g. neutral gas distribution, star formation) in comparison to us and Magrini et al. (2016).

### 3.2 Predicted radial gradients

In Fig. 8, bottom panel, the slopes of the N/O radial gradients [$grad$(N/O)] versus the redshift ($z$) predicted by the CEMs by Mollá & Díaz (2005) are shown. Points for $z = 0$ are obtained from the fittings shown in Fig. 1. Points for $z > 0$ are obtained by carrying out the fitting of Eq. 1 to the CEM predictions (not shown) following the same procedure than for $z = 0$ (see Sect. 2.2). The error bars indicate the uncertainties in the linear fittings to the CEM N/O gradients for a given $z$ value. We can see that $grad$(N/O) flattens out as the $z$ increases showing negative values from $z = 0$ to $z \sim 6$ ($T \sim 0.9$ Gyr), depending on the galaxy mass. For $z \gtrsim 6$, all the gradients become almost flat, until an inversion (positive slopes) for $z \gtrsim 8$.

Unfortunately, abundance gradients of galaxies at high redshift have been rarely derived and only for O/H, mainly due to instrumental constraints to observe the red optical spectral range and/or separate [N II] lines from H$\alpha$, making almost unknown the $grad$(N/O)-$z$ relation at $z > 1$. In fact, Simons et al. (2021) derived the O/H (or $Z$) gradients, relying on strong-line methods, for a sample of 238 star-forming galaxies at $0.6 < z < 2.6$. These authors found that a large fraction of galaxies have flat and positive O/H gradients (see also Sharda et al. 2021; Wang et al. 2022). Moreover, even in the local universe ($z < 1$), flat and positive O/H gradients are derived in some objects (e.g. Krabbe et al. 2008, 2011; Carton et al. 2018; do Nascimento et al. 2022; Boardman et al. 2023), being this result due to, for instance, interactions of galaxies (e.g. Rosa et al. 2014), metal-poor gas accretion (e.g. Kereš et al. 2005) and metal-enriched of the circumgalactic medium by stellar feedback and later gas re-accretion (e.g. Tumlinson et al. 2017). In any case, N/H gradients trend to be steeper than O/H gradients (e.g. Pilyugin et al. 2004), resulting in negative N/O gradients in practically all spiral galaxies in the local universe. Belfiore et al. (2017), by using strong-line methods, derived the O/H and N/O radial gradients for a sample of 550 nearby ($z < 0.15$) galaxies. These authors found that N/O gradients do not flatten in the innermost regions of galaxies, where a flattening of the oxygen abundance gradient is observed in some objects. Thus, probably, future direct estimates of N/O gradients in high-$z$ galaxies could confirm our findings, i.e. N/O gradients in inner disk of spiral galaxies only flatten at very high redshift values ($z \gtrsim 5$).

In Fig. 8, top panel, the logarithm of $(N/O)_0$ versus the redshift predicted by the CEM fittings are shown. The hatched area represents the log(N/O) value for which the nitrogen has mainly a primary origin, i.e. the N/O plateau [log(N/O) = −1.41 ± 0.09] as derived by Berg et al. (2019) through the $T_e$-method and by using observational optical spectroscopic data of local dwarf galaxies. It can be seen that for $z \gtrsim 5$, the N/O nuclear abundance variations with the redshift become smaller trending the $(N/O)_0$ values to be in the shadowed area. Therefore, the nitrogen ISM enrichment in nuclear regions of spiral galaxies at very high redshift is predominately of primary origin. This result has a deep consequence on $Z$ estimates of AGNs through strong-emission line ratios involving nitrogen lines (e.g. [N II]$\lambda$6584/H$\alpha$, Carvalho et al. 2020), in the sense that, in this case, there is a weak (or nonexistent) dependence of such ratios with $Z$. Thus, for $Z$ estimates of AGNs with $z \gtrsim 5$ it is advisable to use calibrations based on, for instance, the $R_{23}$ =([O II]$\lambda$3727+[O III]$\lambda$4959 + $\lambda$5007)/H$\beta$ index, as suggested by Dors (2021).

Up to now, due to the lack of N/O estimates for AGNs at high-$z$ in the literature, it is only possible to compare our N/O nuclear abundance predictions with those derived from integrated spectra of SFs. In this regard, in Fig. 9 our $(N/O)_0$ predictions are compared with abundances for high redshift SFs derived through the $T_e$-method and compiled from the literature. Moreover, it is worth mentioning that we are comparing abundance estimates in distinct object classes, which have different star formation histories (see Riffel et al. 2023). Taking into account the high observational uncertainties in the N/O direct estimations (~ 0.2 dex) of the SFs and the different nature of AGNs and SFs, we are able to assume that most part (9/12) of the N/O direct estimates are reproduced by our CEM predictions.

The $(N/O)_0 - z$ relation obtained from our nuclear abundance predictions is represented by:

$$\log(N/O)_0 = (a \times z^2) + (b \times z) + c, \quad (10)$$

where a = 0.007 ± 0.001, b = −0.13 ± 0.01 and c = −0.83 ± 0.02. This relation is represented in Fig. 9 by a black curve.

In Fig. 10, bottom panel, our predicted CEM O/H values for the nuclear galaxy regions are plotted against the redshift. The expected tendency of the decrement of O/H (or $Z$) with the redshift is easily noticeable. This relation can be represented by a linear regression given by:

$$12 + \log(O/H)_{CEM} = -(0.24 \pm 0.01 \times z) + 8.64 \pm 0.06 \quad (11)$$

and it is plotted as a black line. With the goal of comparing our results, we overlapped on this figure other estimates taken from the literature. In the bottom panel of Fig. 9, our O/H abundance predictions as a function of $z$ are compared with:

- Estimates of O/H abundances obtained by using the $T_e$-method for 48 SFs at the redshift range $1.4 \lesssim z \lesssim 9.5$. These estimates are represented by pink points while the pink line is the liner fitting to them given by

$$12 + \log(O/H)_{SF} = -0.03 \pm 0.02 \times z + 7.89 \pm 0.11. \quad (12)$$

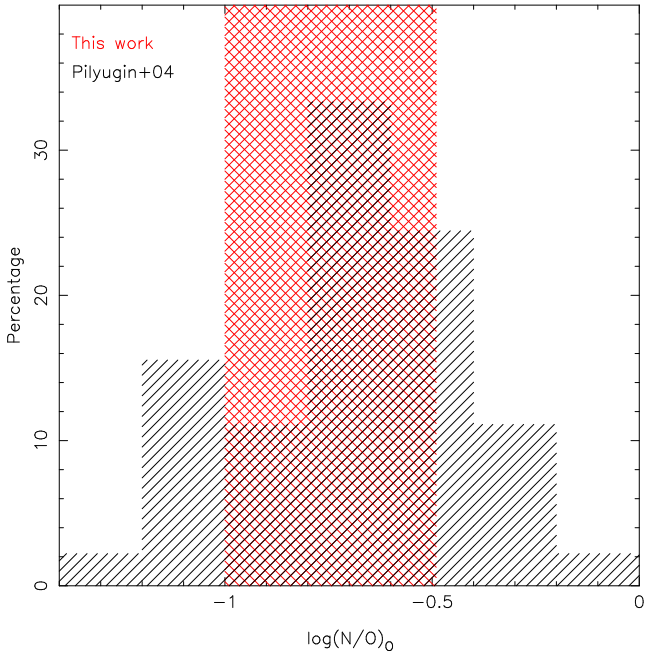

**Figure 7.** Distribution (in black) of the central intercept N/O abundance [log(N/O)$_0$] values derived by Pilyugin et al. (2004) using the *P*-method (Pilyugin 2001) for 45 local ($z < 0.02$) galaxies. Red hatched area represents the log(N/O)$_0$ range derived from our CEM fittings to direct radial gradients of four local spirals (see Table 1).

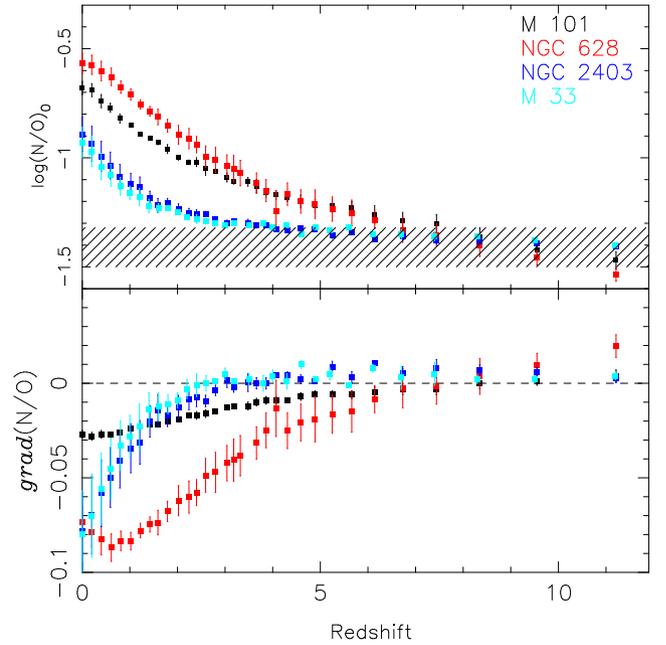

**Figure 8.** Chemical evolution model predictions of the radial gradient slope (in units of kpc dex$^{-1}$) for N/O (bottom panel) and for the logarithm of (N/O)$_0$ (top panel). Distinct colors indicate the predictions of CEMs by Mollá & Díaz (2005) for the spiral galaxies as indicated. The dashed line in bottom panel represents the zero value of the N/O gradient. Hatched area in top panel represents the N/O plateau [log(N/O) = $-1.41 \pm 0.09$] derived by Berg et al. (2019).

- O/H abundances for SF galaxies at the redshift range $0.07 \lesssim z \lesssim 3.5$, with stellar masses ranging from $9.0 \lesssim (\log(M_*/M_\odot) \lesssim 11.0$ derived through strong-line methods by Maiolino et al. (2008) are included in the figure as a region delimited by orange lines.

- Estimates (blue points) for narrow line regions of AGNs ($1.2 \lesssim z \lesssim 3.8$) based on strong emission lines observed in the ultraviolet wavelength range and obtained from comparison with photoionization model predictions by Dors et al. (2019).

In Fig. 9, top panel, our O/H abundance predictions as a function of $z$ are compared with:

- Estimates (pink points) obtained through absorption lines for 151 Damped L$\alpha$ and sub-Damped L$\alpha$ galaxies (DLA) at a redshift range of $0.2 \lesssim z \lesssim 5.1$. A linear fit to these estimates yields:

$$12 + \log(O/H)_{DLA} = -(0.37 \pm 0.03) \times z + (8.33 \pm 0.09). \quad (13)$$

The metallicity ($Z$) estimates for AGNs by Dors et al. (2019) and those for DLAs were converted into O/H abundances according to the relation:

$$12 + \log(O/H) = 12 + \log[(Z/Z_\odot) \times 10^{\log(O/H)_\odot}], \quad (14)$$

being log(O/H)$_\odot = -3.31$ the solar oxygen abundance (Allende Prieto et al. 2001). It is worth noting that, for $z \lesssim 5$, our predicted nuclear O/H abundances are in consonance with those for SFs and AGNs. Only a few active nuclei show O/H values higher, by $\sim 0.2$ dex, than our CEM estimates. However, for $z \gtrsim 5$, our predictions yield O/H values systematically lower than those via the $T_e$-method for SFs, showing the latter a small O/H abundance decrease with $z$. The origin of this discrepancy is likely due to an observational bias, in the sense that the galaxies with higher luminosity (and metallicity) are easier to observe (e.g. Nagao et al. 2006c). The DLA estimates show lower O/H values and a steeper decrease with $z$ than our predictions. This behavior was also found by Dors et al. (2014), who compared $Z$ estimates of AGNs and DLAs, and it is probably due to the use of distinct methods (i.e. $Z$ estimates via emission and absorption lines) rather than an observational bias.

In summary, from the comparisons presented in Figs. 9 and 10, we are able to conclude that our nuclear N/O and O/H abundances are in consonance with those for the most part of SFs located at $z \lesssim 5$. For $z \gtrsim 5$, we found a discrepancy that could be due to the existence of a bias in galaxy observations. In the case of DLAs the disagreement could be attributed to the distinct methods applied to estimate $Z$.

Finally, in Fig. 11, our CEM oxygen abundance predictions [(O/H)$_0$] are compared with the extrapolated O/H abundances to the nuclear regions ($R = 0$ kpc) of 14 high-$z$ galaxies observed by Simons et al. (2021). These galaxies have masses in the $8.6 \lesssim \log(M_*/M_\odot) \lesssim 10$ range, redshifts in the $1 \lesssim z \lesssim 2$ range and O/H radial gradients derived through the izi Bayesian photoion-ization fitting code (Blanc et al. 2015). It can be seen in Fig. 11 that our CEM estimates for nuclear regions of galaxies are in most part in agreement with those from Simons et al. (2021). Again, more estimations of nuclear metallicities or radial abundance gradients for high-$z$ ($z > 2$) galaxies are necessary to produce a better validation of our estimates.

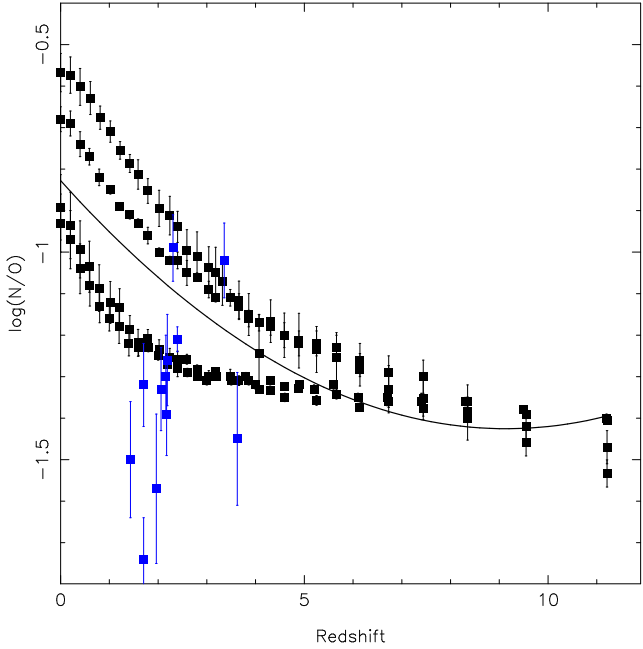

**Figure 9.** Left panel: Logarithm of the N/O abundance ratio versus the redshift. Blue points represent N/O abundances via the $T_e$-method for SFs taken from the literature and obtained by Kojima et al. (2017), Christensen et al. (2012a,b), Steidel et al. (2014), Erb et al. (2016), Fosbury et al. (2003), Villar-Martín et al. (2004), Yuan & Kewley (2009), Bayliss et al. (2014), Pettini et al. (2010), James et al. (2014), Stark et al. (2014), and Steidel et al. (2016). Black points represent our CEM (N/O)$_0$ estimates listed in Tables A1-A4. The blue curve represents the fitting to our estimates (Eq. 10).

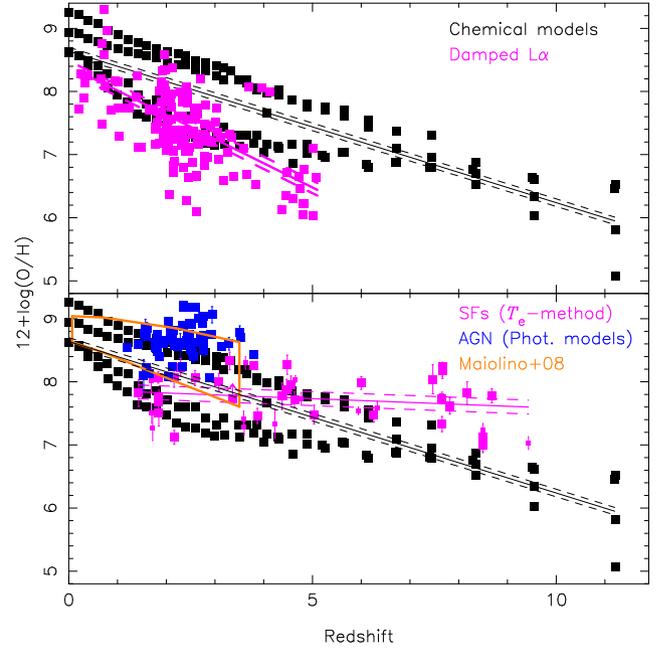

**Figure 10.** Logarithm of the O/H abundance versus the redshift. Bottom panel: Black points represent O/H abundances as a function of redshift obtained from the N/O values (shown in left panel) and applying the (N/O)-(O/H) relation given by Eq. 7. Pink points represent direct estimates for SFs by Clarke et al. (2023), Sanders et al. (2023b); Arellano-Córdova et al. (2022); Sanders et al. (2016b), Villar-Martín et al. (2004), Yuan & Kewley (2009), Brammer et al. (2012), Christensen et al. (2012b), James et al. (2014), Bayliss et al. (2014), Kojima et al. (2017), Gburek et al. (2019), Curti et al. (2023b), Citro et al. (2023), and Laseter et al. (2023). Blue points: AGN metallicity estimates via photoionization models by Dors et al. (2019). Orange lines delimit the region occupied by SF galaxies [$9.0 \lesssim \log(M_*/M_\odot) \lesssim 11.0$] whose O/H abundances were derived through strong-line methods by Maiolino et al. (2008). Top panel: Pink points represent metallicity estimations for Damped L$\alpha$ and sub-Damped L$\alpha$ galaxies via absorption lines by Rafelski et al. (2013), Fox et al. (2007) and Kulkarni et al. (2005). Black points are as in bottom panel.

It is worth mentioning that, in general, supersolar metallicities are derived for Broad Line Regions of AGNs (e.g. Dietrich et al. 2003; Nagao et al. 2006a; Juarez et al. 2009; Sameshima et al. 2017), reaching up $\sim 18 Z_\odot$ in N-loud quasi-stellar objects (QSOs, e.g. Batra & Baldwin 2014). Recent metallicity estimates for high-redshift ($5.8 < z < 7.5$) quasars have also found high metallicities ($\sim 2-4Z_\odot$, Lai et al. 2022), indicating no evolution with the redshift (e.g. Dors et al. 2014; Onoue et al. 2020). As shown by Dors et al. (2019), $Z$ values in BLRs are a factor of about 2-3 higher than those in NLRs and are inconsistent with predictions of chemical evolution models (see Fig. 11). The origin of this discrepancy has been attributed to star formation in accretion disks (e.g. Fan & Wu 2023; Huang et al. 2023; Cheng & Loeb 2023; Wang et al. 2023; Dittmann et al. 2023), uncertainties in BLR metallicity estimates from some UV lines (e.g. Matsuoka et al. 2017) and the fact that broad lines originate in a small region with a radius lower than 1 pc (e.g. Kaspi et al. 2000; Suganuma et al. 2006), which may evolve more rapidly than the NLRs (e.g. Matsuoka et al. 2018). It is beyond of the scope of this study to investigate the BLR/NLR metallicity discrepancy. We only pointed out that, apparently, NLR estimates seem to be better suited as a proxy for the properties of the host galaxy since the spatial extent of the NLR region ($10^{1-4}$ pc) is larger than the BLR ($< 1$ pc, Armah et al. 2023).

### 3.3 Optical diagnostic diagrams

The main goal of this work is to analyze which are the physical parameters (i.e. O/H and $U$), and the correspondent $z$ values, that let AGNs fall below the maximum starburst lines in diagnostic diagrams. In this regard, in Figs. 12 and 13, the [O III]/H$\beta$ versus [O I]/H$\alpha$, [N II]/H$\alpha$ and [S II]/H$\alpha$ BPT diagrams containing our photoionization model results are shown. Also in these figures, the He II/H$\beta$ versus [N II]/H$\alpha$ diagram, proposed by Shirazi & Brinchmann (2012) to separate AGNs from SFs, are shown. In Fig. 12, the colour bars represent the O/H abundance of the photoionization models, while in Fig. 13 the corresponding redshifts for which the theoretical AGNs reach the O/H abundance. The model results assuming different log $U$ values are represented by distinct symbols, as indicated. In each diagram the maximum starburst lines, taken from the literature, are also plotted. Likewise, lines to separate AGNs from low-ionization nuclear emission-line regions (LINERs, represented by, simplicity, by LIN) are plotted in the [O III]/H$\beta$ versus [N II]/H$\alpha$ and [S II]/H$\alpha$ diagrams. Also in Fig. 12 and Fig. 13, observational emission line ratio intensities, corrected by reddening and after apply stellar population continuum subtraction, of galaxies taken from Sloan Digital

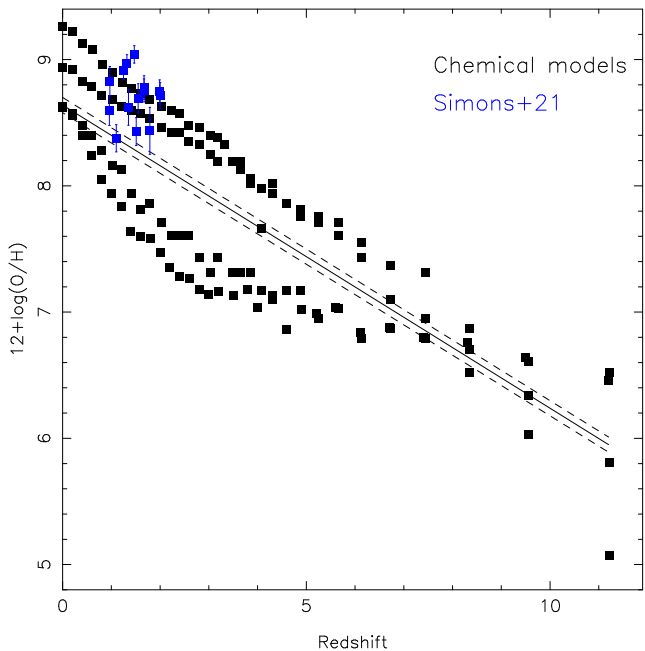

**Figure 11.** Such as Fig. 10 but with the blue points representing the extrapolation of the O/H radial gradients to the nuclear regions (galactocentric distance equal to 0) by Simons et al. (2021) for 14 galaxies ($8.6 \lesssim \log(M_*/M_\odot) \lesssim 10$, $1 \lesssim z \lesssim 2$). The O/H gradients were calculated by these authors through the IZI Bayesian photoionization fitting code (Blanc et al. 2015).

Sky Survey (SDSS) DR17[6] (Abdurro'uf et al. 2022) database are shown. The procedures applied to these data are the same as in Dors et al. (2020).

Firstly, we can see in Fig. 12 that AGN models with $12 + \log(O/H) \gtrsim 9.0$ [$(Z/Z_\odot) \gtrsim 2.0)$], independently of the assumed $U$ value, are located near or just below the maximum starburst lines of the classical BPT diagrams (see also Feltre et al. 2016; Nakajima & Maiolino 2022). However, similar very high oxygen abundance (or metallicity) is obtained in some few AGNs in the local universe ($z < 1$). In fact, O/H estimates for a large sample (463 objects) of local ($z < 0.4$) AGNs obtained through strong-line methods and inferred extrapolating radial abundance gradients by Dors et al. (2020) showed that objects with $12 + \log(O/H) \gtrsim 9.0$ comprehends only ~ 10 per cent of their sample. In any case, strong-line methods based on [N II]/[O II] and [N II]/H$\alpha$ lines ratios proposed by Castro et al. (2017) and Carvalho et al. (2020), respectively, can be used to distinguish between metal-rich and metal-poor AGNs.

Secondly, in Fig. 12, it can be seen the known result (see e.g. Feltre et al. 2016) that low metallicity AGNs occupy a large area in the BPT diagrams where SFs are located. In each diagram, we indicated the average O/H abundance for which AGNs pass through the maximum starburst line, i.e. AGNs with $12 + \log(O/H) \lesssim 8.0$ or $(Z/Z_\odot) \lesssim 0.2$ are located in BPT regions occupied by SFs. In each diagram of Fig. 13 we indicated the average redshift, according to our CEM results, for which AGNs reach the oxygen abundance of $12+\log(O/H) \approx 8.0$.

The results obtained by Hirschmann et al. (2022), who assumed as input parameters metallicity results from ILLUSTRISTNG simulations, indicate that:

[6] SDSS DR17 spectroscopic data are available at `https://dr17.sdss.org/optical/plate/search`.

- AGNs with $12+\log(O/H) \lesssim 8.4$ [$(Z/Z_\odot) \lesssim 0.5$] move towards the maximum starburst line.
- AGN galaxies with these abundances are preferably found at $z \gtrsim 1$.

Our analysis indicates that:

- AGNs with $12 + \log(O/H) \lesssim 8.0$ [$(Z/Z_\odot) \lesssim 0.2$] occupy the SF zone in standard BPT diagrams.
- AGNs reach this abundance at $z \sim 3.7$.

The abundance discrepancy between our results and those obtained by Hirschmann et al. (2022) is, probably, due to distinct photoionization parameters [e.g. SED, (N/O)-(O/H) relation, dust content, atomic parameters] assumed in both works. In any case, assuming the abundance uncertainty of ~ 0.2 dex for strong-line methods (e.g. Denicoló et al. 2002), the O/H values derived by us and by Hirschmann et al. (2022) are in agreement to each other.

Regarding the redshift value discrepancy, it could be due to the different $Z - z$ relations derived in both works. Unfortunately, this relation is not presented by Hirschmann et al. (2022). However, CEM predictions by Hirschmann et al. (2017), relied on the SPHGAL code (Hu et al. 2014) for nuclear regions of three hypothetical galaxies with $\log(M_{gal}/M_*) \sim 11$, indicate that galaxies with redshift around $z = 4$ reach abundances that result in emission-line ratios in the SF-zone, in good agreement with our present estimation.

We compare our $Z - z$ relation with others derived by considering some CEMs available in the literature. For that, we consider:

- Chemical Evolution models by Pei et al. (1999) calibrated by using optical imaging surveys of quasars and the COBE DIRBE and FIRAS extragalactic infrared background measurements. These models are valid for $0.0 \leq z \leq 5.0$.
- Semi-analytic models of galaxy formation set within the cold dark matter merging hierarchy built by Somerville et al. (2001). These models are able to reproduce the metallicity of high-redshift Lyman-break galaxies in the redshift range $z = 0.0 - 4.0$.
- Metallicity predictions from ILLUSTRISTNG simulations by Torrey et al. (2019) for the redshift range $z = 0.0 - 10.0$ and for hypothetical galaxies with masses in the range $8.0 \leq \log(M/M_*) \leq 10.5$. We consider the average O/H value predicted by this mass interval. Although Hirschmann et al. (2022) used the predictions of these models, these authors assumed as input in their photoionization models metallicity values obtained in a co-moving sphere of 1 kpc radius around the hypothetical nuclear galaxy region, about the same approach that the one assumed in the present work.
- Recent multi-zone galactic chemical evolution models proposed by Johnson et al. (2023) which reproduces the (N/O)-(O/H) relation found for the Milky Way stars and for the gas phase of external galaxies. We selected CEMs by Johnson et al. (2023) considering radial stellar migration proceeds and the O/H abundance predicted by the innermost galactocentric distance $R$, i.e $R = 4$ kpc. This procedure was adopted in order to obtain the nearest nuclear abundances for galaxies with distinct evolutionary stages. These models are appropriate for Milky Way mass galaxies with $10.5 \leq \log(M_{gal}/M_*) \leq 11$.

Except for Johnson et al. (2023), the CEMs listed above predict an average metallicity for the ISM for the whole galaxy at a given redshift, i.e. they do not predict nuclear metallicity estimates in the hypothetical galaxies. In Fig. 14, the O/H abundance as a function of the redshift predicted by the CEMs listed above is shown. Also in this figure, the redshift at which AGNs reach the abundance $12+\log(O/H)=8.0$ [$(Z/Z_\odot) = 0.2$] are indicated for the distinct CEMs (green points). Hereafter, we define this redshift value, i.e. the redshift in which AGNs reach 12+log(O/H)=8.0, as the $z_{\rm AGN}$-limit. We can see that the $z_{\rm AGN}$-limit is strongly dependent on the CEMs considered, ranging from $z \sim 1.4$ to $z \sim 10$ for all the models except the model by Johnson et al. (2023). Since the nuclear galaxy abundance trends to be higher than the one of the whole galaxy (e.g. Hirschmann et al. 2017) or than that of the ISM, the $z_{\rm AGN}$-limit indicated in Fig. 14 must be interpreted as a lower limit. In this scenario, our result that AGNs with $12 + \log({\rm O/H}) \lesssim 8.0$ are preferable found at $z \gtrsim 3.7$ is in consonance with the estimated range for three of the CEMs from the literature. The CEM results from Johnson et al. (2023) predict higher O/H abundances along the redshift interval considered. Possibly, it is due to the models proposed by Johnson et al. (2023) being designed for massive galaxies (see above).

Finally, we discuss our results for the He II/H$\beta$ versus [N II]/H$\alpha$ diagram shown in the lower right panels of Figs. 12 and 13. As was mentioned before, this diagram was suggested by Shirazi & Brinchmann (2012) to separate AGNs from SFs. These authors defined the maximum starburst line by using observational data of galaxies ($z \lesssim 0.2$) from the Sloan Digital Sky Survey (SDSS) DR7 (York et al. 2000; Stoughton et al. 2002) previously classified through classical BPT diagrams. Since the nebular He II emission has a high ionization potential (54.4 eV), it is preferably produced by a hard ionizing spectrum, which may indicate AGN activity. Recently, Tozzi et al. (2023), who used observational data from Mapping Nearby Galaxies at APO survey (MaNGA DR15, Bundy et al. 2015), found an overall increased fraction (2 per cent) of AGNs in MaNGA galaxies when the He II diagram is used (11 per cent) in comparison to the BPT-only census (9 per cent). It is worth to mention that the He II/H$\beta$ versus [N II]/H$\alpha$ diagram suffers a limitation due to the difficulty to measure the He II $\lambda4685$ emission line (about 4-10 times weaker than H$\beta$, see Fig. 12). In fact, Shirazi & Brinchmann (2012), based on SDSS spectroscopic data ($z \lesssim 0.4$), showed that the He II $\lambda4685$ is most frequently measured in SF galaxies at low redshift ($z \lesssim 0.1$, see also Sartori et al. 2015; Bär et al. 2017; Koss et al. 2017; Wang & Kron 2020; Mayya et al. 2020; Oh et al. 2022; Tozzi et al. 2023). However, recent observations with the JWST (e.g. Übler et al. 2023) have shown the reliability of measuring this line in AGNs at very high redshift ($z \gtrsim 5$). Thus, the new generation of extreme large and space telescopes will increase the feasibility of diagnostic diagrams using He lines, such as the He II/H$\beta$ versus [N II]/H$\alpha$ (Shirazi & Brinchmann 2012) and [O III]/[O II] versus He II/He I (Dors et al. 2022).

We can see in Figs. 12 and 13 that our cosmological predictions indicate that, independently of the metallicity and of the redshift, AGNs are always above the maximum starburst line in the He II/H$\beta$ versus [N II]/H$\alpha$ diagram, indicating that this diagram is the most appropriated to classify objects at any redshift, independently of the AGN nebular parameters (i.e. O/H and $U$). It is worth to mention that, as can be seen in Figs. 12 and 13, our theoretical results indicate a similar AGN-SF separation line in agreement with that proposed by Molina et al. (2021), who presented a sample of 81 dwarf galaxies ($M_* \leq 3 \times 10^9 M_\odot$) with detectable [Fe x]$\lambda6374$ coronal emission line, indicating accretion onto massive black holes. The criterion proposed by Molina et al. (2021) suggests that objects with

$$\log({\rm He\ II}\lambda4686/{\rm H}\beta) \gtrsim -1.0, \quad (15)$$

are classified as AGNs, otherwise, SFs. In Figs. 12 and 13 one can see that the photoionization model results do not reproduce some SDSS observational data. This is due to the limited sample of galaxies considered by us (which define the space of model parameters) in comparison to the data from SDSS sample. It is worth mentioning that, for a better match, is necessary a larger sample of spiral galaxies with direct radial N/O gradients.

## 4 CONCLUSIONS

We investigated the reliability of optical diagnostic diagrams, based on emission-line ratios [$4000 < \lambda({\rm \AA}) < 7000$], in the framework of classifying Active Galactic Nuclei (AGNs) according to the metallicity evolution in the redshift range $0 \leq z \leq 11.2$. With this aim, we fit results of Chemical Evolution Models (CEMs) to the radial abundance gradients derived through direct estimates of electron temperatures ($T_e$-method) in a sample of four local spiral galaxies. Unlike the majority of previous studies, we consider as metallicity tracer the N/O abundance ratio to select the representative CEMs for each object belonging to our galaxy sample. The N/O abundance radial predictions provided by the CEMs were extrapolated to the nuclear regions (galactocentric distance $R = 0$ kpc) in order to infer N/O and O/H abundances in theoretical galaxies at distinct evolutionary stages. These resulting abundance ratios were used as input parameters in AGN photoionization models built with the Cloudy code. Extensive comparison between CEM predictions and direct abundance estimates, in a wide range of redshift, was performed in order to validate our cosmic abundance values. From our analysis we conclude that:

- BPT diagnostic diagrams: i.e. [O III]$\lambda5007$/H$\beta$ versus [N II]$\lambda6584$/H$\alpha$, [O I]$\lambda6300$/H$\alpha$ and [S I]($\lambda6716 + \lambda6731$)/H$\alpha$ are able to classify AGNs with oxygen abundances $12 + \log({\rm O/H}) \gtrsim 8.0$ [$(Z/Z_\odot) \gtrsim 0.2$].
- Our CEM predictions show that AGNs reach the oxygen abundance of $12 + \log({\rm O/H}) = 8.0$ preferably at redshift $z \sim 4$, indicating that BPT diagrams break down for higher redshift values.
- We found that the He II$\lambda4685$/H$\beta$ versus [N II]$\lambda6584$/H$\alpha$ diagram is able to separate AGNs from star-forming regions in the redshift range $0 \lesssim z \lesssim 11.2$.

## ACKNOWLEDGEMENTS

We are grateful to the anonymous referee for her/his very useful comments and suggestions that helped us to clarify and improve this work. OLD is grateful to Fundacão de Amparo à Pesquisa do Estado de São Paulo (FAPESP) and Conselho Nacional de Desenvolvimento Científico e Tecnológico (CNPq). G.S.I. acknowledges financial support from FAPESP (Fundação de Amparo à Pesquisa do Estado de São Paulo, Proj. 2022/11799-9).

## DATA AVAILABILITY

The data underlying this article will be shared on reasonable request to the corresponding author.

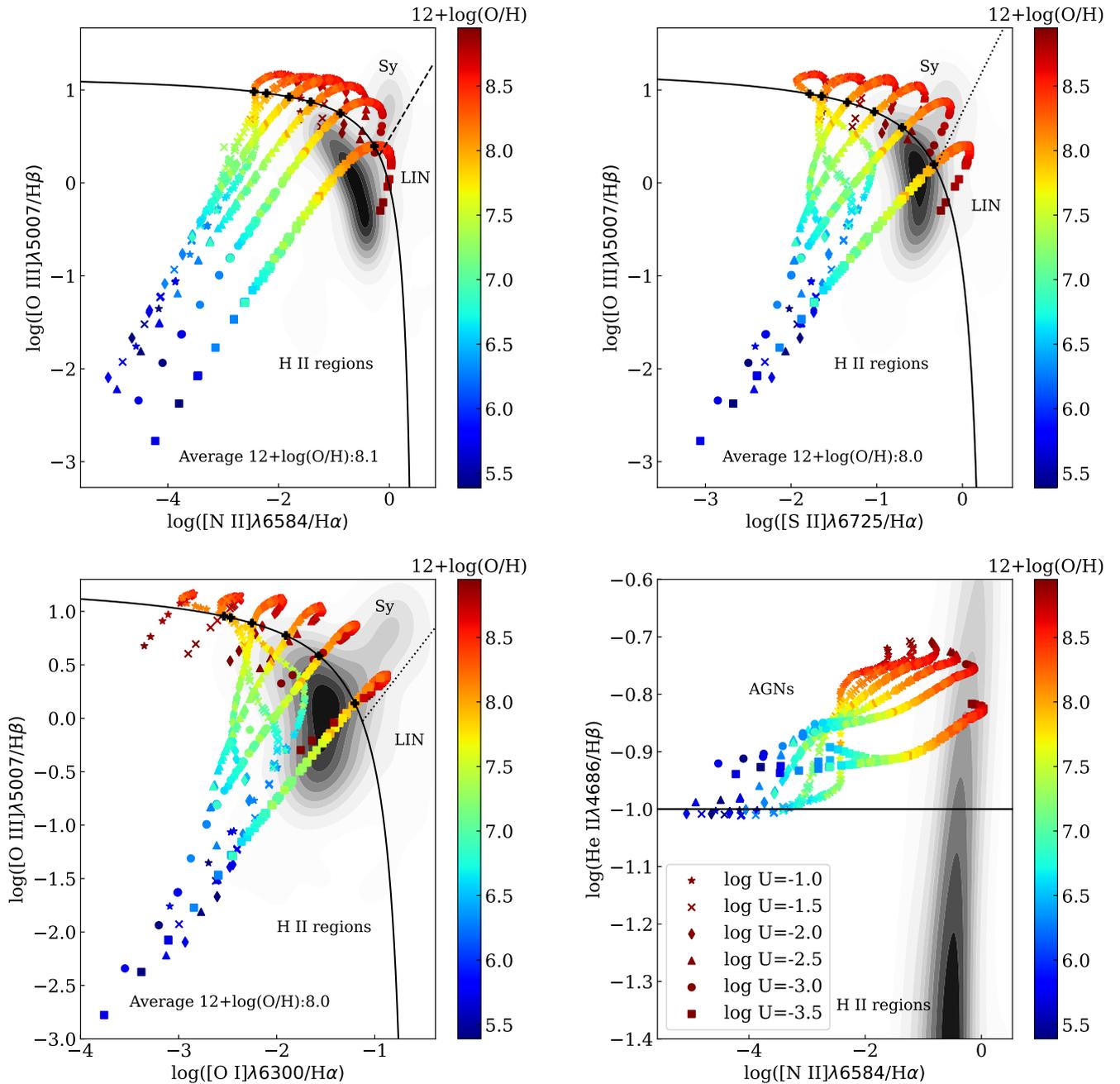

**Figure 12.** Diagnostic diagrams involving optical emission-line ratios as indicated. In the [O III]/Hβ versus [O I]/Hα, [N II]/Hα and [S II]/Hα the black curves represent the maximum starburst line proposed by Kewley et al. (2001). The dashed line separates AGNs from LINERs (refereed as LIN) as suggested by Kewley et al. (2006). The [S II]λ6725 corresponds to the sum [S II]λ6716 + λ6731. In the He II/Hβ versus [N II]/Hα diagram, the line represents the maximum starburst line proposed by Molina et al. (2021). In all diagrams, points represent results of our photoionization models (see Sect. 2.3) by using as input parameters the abundances predicted by CEMs (see Sect. 2.2) of Mollá & Díaz (2005). Symbols represent photoionization model with iso-$U$, as indicated. Color bars indicate the oxygen abundance assumed in the photoionization models. AGNs with oxygen abundance values lower than the ones indicated in each of the three BPT diagrams [average 12+log(O/H) values] are located below the maximum starburst lines. Contours are galaxies whose data were taken from SDSS DR-17 (Abdurro'uf et al. 2022).

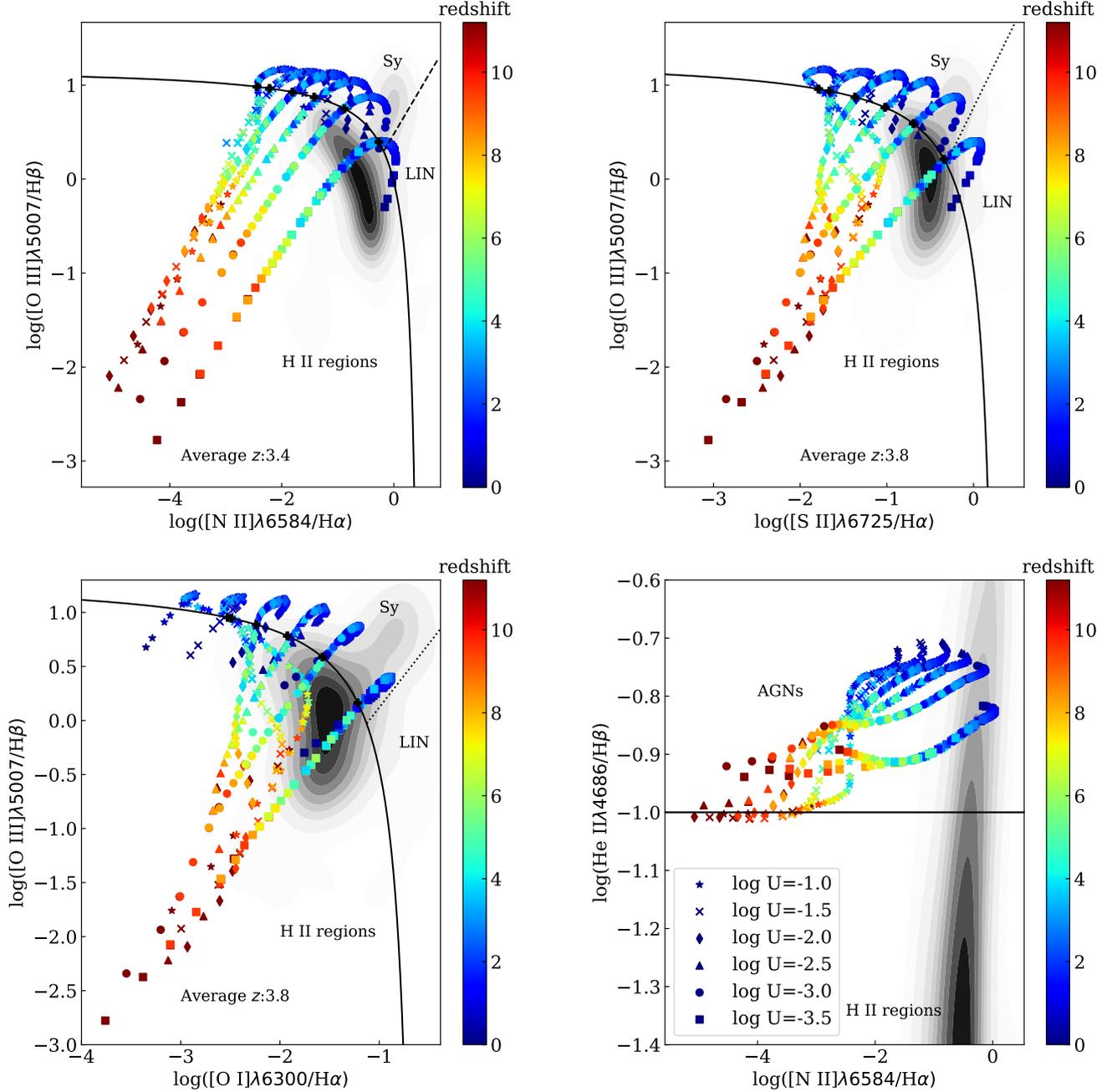

**Figure 13.** As the Fig. 10 but with the color bars indicating the redshift value of each photoionization model representing an AGN. AGNs with redshift values ($z_{AGN}$-limit) lower than the ones indicated in each BPT diagram (average z values) are located below the maximum starburst lines.

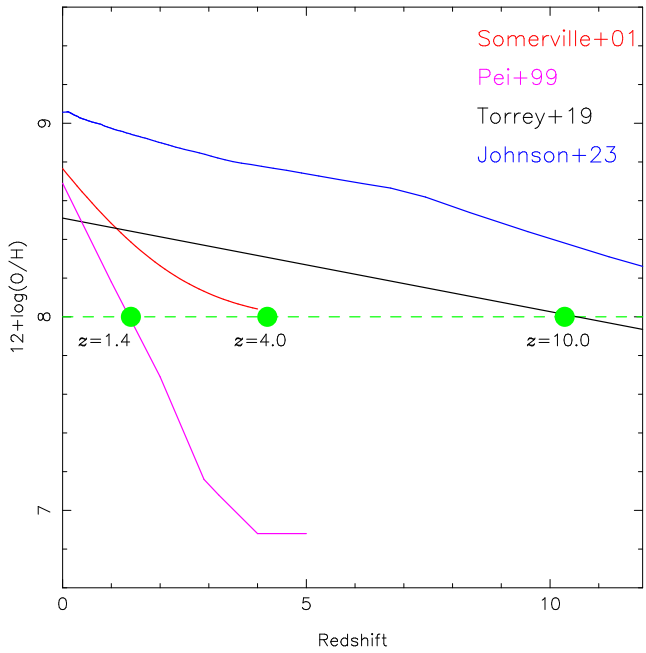

**Figure 14.** Oxygen abundances [in units of 12+log(O/H)] versus the redshift predicted by the CEMs of Somerville et al. (2001), Pei et al. (1999), Torrey et al. (2019), and Johnson et al. (2023), as indicated. AGNs with lower values than 12+log(O/H)=8.0, indicated by a horizontal line, are located below the maximum starburst lines in BPT diagrams, as shown in Fig. 12. Green points represent the interception between the horizontal line and the chemical models. The redshift value ($z_{\rm AGN}$-limit) for which the chemical models predict the 12+log(O/H)=8.0 abundance are indicated.

# APPENDIX A: TABLES

This paper has been typeset from a T$_{\rm E}$X/L$^{\rm A}$T$_{\rm E}$X file prepared by the author.

**Table A1.** Slope and central intersect values of the N/O regressions estimated using the predictions of the CEMs built by Mollá & Díaz (2005) for different redshift. The term *grad* represents the slope of the gradients in units of dex kpc$^{-1}$. The term log(N/O)$_0$ represents the extrapolated abundance ratio to the galactocentric distance ($R$) equal to zero. The central oxygen abundance 12+log(O/H)$_0$ is estimated using log(N/O)$_0$ and Eq.7.

| Redshift | *grad* (dex kpc$^{-1}$) | M 101 log(N/O)$_0$ | 12+log(O/H)$_0$ |
|---|---|---|---|
| 0.00 | −0.027 ± 0.002 | −0.68 ± 0.03 | 8.94 |
| 0.20 | −0.028 ± 0.002 | −0.69 ± 0.03 | 8.92 |
| 0.40 | −0.027 ± 0.002 | −0.74 ± 0.03 | 8.83 |
| 0.59 | −0.027 ± 0.001 | −0.77 ± 0.02 | 8.79 |
| 0.79 | −0.026 ± 0.001 | −0.82 ± 0.02 | 8.72 |
| 1.02 | −0.024 ± 0.001 | −0.85 ± 0.01 | 8.68 |
| 1.21 | −0.023 ± 0.001 | −0.89 ± 0.00 | 8.63 |
| 1.42 | −0.022 ± 0.001 | −0.91 ± 0.01 | 8.60 |
| 1.59 | −0.022 ± 0.001 | −0.93 ± 0.01 | 8.57 |
| 1.79 | −0.020 ± 0.001 | −0.96 ± 0.02 | 8.53 |
| 2.02 | −0.019 ± 0.001 | −1.00 ± 0.01 | 8.46 |
| 2.24 | −0.017 ± 0.001 | −1.02 ± 0.01 | 8.42 |
| 2.40 | −0.017 ± 0.002 | −1.02 ± 0.02 | 8.42 |
| 2.59 | −0.016 ± 0.002 | −1.05 ± 0.03 | 8.35 |
| 2.80 | −0.015 ± 0.001 | −1.06 ± 0.01 | 8.33 |
| 3.04 | −0.013 ± 0.001 | −1.09 ± 0.02 | 8.25 |
| 3.18 | −0.012 ± 0.001 | −1.11 ± 0.01 | 8.19 |
| 3.48 | −0.012 ± 0.002 | −1.11 ± 0.02 | 8.19 |
| 3.66 | −0.010 ± 0.002 | −1.13 ± 0.03 | 8.13 |
| 3.86 | −0.009 ± 0.002 | −1.16 ± 0.02 | 8.02 |
| 4.07 | −0.009 ± 0.025 | −1.17 ± 0.02 | 7.98 |
| 4.31 | −0.009 ± 0.001 | −1.18 ± 0.02 | 7.94 |
| 4.59 | −0.007 ± 0.002 | −1.20 ± 0.03 | 7.86 |
| 4.89 | −0.006 ± 0.002 | −1.22 ± 0.03 | 7.76 |
| 5.25 | −0.006 ± 0.002 | −1.22 ± 0.03 | 7.76 |
| 5.66 | −0.006 ± 0.002 | −1.23 ± 0.03 | 7.71 |
| 6.14 | −0.005 ± 0.003 | −1.26 ± 0.04 | 7.55 |
| 6.73 | −0.003 ± 0.003 | −1.29 ± 0.04 | 7.37 |
| 7.44 | −0.003 ± 0.003 | −1.30 ± 0.04 | 7.31 |
| 8.35 | −0.000 ± 0.003 | −1.36 ± 0.04 | 6.87 |
| 9.55 | +0.001 ± 0.002 | −1.42 ± 0.03 | 6.34 |
| 11.21 | +0.004 ± 0.003 | −1.47 ± 0.04 | 5.81 |

**Table A2.**

| | NGC 628 | | |
|---|---|---|---|
| Redshift | $grad$ (dex kpc$^{-1}$) | $\log(\text{N/O})_0$ | $12+\log(\text{O/H})_0$ |
| 0.00 | $-0.0735 \pm 0.008$ | $-0.56 \pm 0.04$ | 9.26 |
| 0.20 | $-0.0785 \pm 0.007$ | $-0.57 \pm 0.04$ | 9.22 |
| 0.40 | $-0.0826 \pm 0.008$ | $-0.60 \pm 0.04$ | 9.13 |
| 0.61 | $-0.0869 \pm 0.007$ | $-0.62 \pm 0.03$ | 9.08 |
| 0.81 | $-0.0833 \pm 0.005$ | $-0.67 \pm 0.02$ | 8.96 |
| 1.02 | $-0.0834 \pm 0.004$ | $-0.70 \pm 0.02$ | 8.90 |
| 1.22 | $-0.0779 \pm 0.003$ | $-0.75 \pm 0.02$ | 8.82 |
| 1.42 | $-0.0746 \pm 0.003$ | $-0.78 \pm 0.02$ | 8.77 |
| 1.59 | $-0.0736 \pm 0.006$ | $-0.81 \pm 0.03$ | 8.73 |
| 1.79 | $-0.0677 \pm 0.005$ | $-0.85 \pm 0.02$ | 8.68 |
| 2.02 | $-0.0620 \pm 0.007$ | $-0.89 \pm 0.04$ | 8.63 |
| 2.24 | $-0.0601 \pm 0.008$ | $-0.91 \pm 0.04$ | 8.60 |
| 2.40 | $-0.0578 \pm 0.006$ | $-0.93 \pm 0.03$ | 8.57 |
| 2.59 | $-0.0487 \pm 0.008$ | $-0.99 \pm 0.04$ | 8.48 |
| 2.80 | $-0.0470 \pm 0.010$ | $-1.00 \pm 0.05$ | 8.46 |
| 3.04 | $-0.0422 \pm 0.009$ | $-1.03 \pm 0.05$ | 8.40 |
| 3.18 | $-0.0406 \pm 0.011$ | $-1.04 \pm 0.06$ | 8.38 |
| 3.32 | $-0.0383 \pm 0.010$ | $-1.06 \pm 0.05$ | 8.33 |
| 3.66 | $-0.0316 \pm 0.008$ | $-1.11 \pm 0.04$ | 8.19 |
| 3.86 | $-0.0249 \pm 0.009$ | $-1.15 \pm 0.05$ | 8.06 |
| 4.07 | $-0.0132 \pm 0.014$ | $-1.24 \pm 0.09$ | 7.66 |
| 4.31 | $-0.0251 \pm 0.009$ | $-1.16 \pm 0.05$ | 8.02 |
| 4.59 | $-0.0207 \pm 0.009$ | $-1.20 \pm 0.05$ | 7.86 |
| 4.89 | $-0.0190 \pm 0.012$ | $-1.21 \pm 0.06$ | 7.81 |
| 5.25 | $-0.0162 \pm 0.010$ | $-1.23 \pm 0.05$ | 7.71 |
| 5.66 | $-0.0150 \pm 0.010$ | $-1.25 \pm 0.05$ | 7.61 |
| 6.14 | $-0.0083 \pm 0.007$ | $-1.28 \pm 0.03$ | 7.43 |
| 6.73 | $-0.0026 \pm 0.010$ | $-1.33 \pm 0.05$ | 7.10 |
| 7.44 | $-0.0017 \pm 0.009$ | $-1.35 \pm 0.05$ | 6.95 |
| 8.35 | $+0.0036 \pm 0.009$ | $-1.40 \pm 0.05$ | 6.52 |
| 9.55 | $+0.0098 \pm 0.005$ | $-1.45 \pm 0.03$ | 6.03 |
| 11.21 | $+0.0196 \pm 0.005$ | $-1.53 \pm 0.03$ | 5.07 |

Table A3.

| | NGC 2403 | | |
|---|---|---|---|
| Redshift | $grad$ (dex kpc$^{-1}$) | $\log(N/O)_0$ | $12+\log(O/H)_0$ |
| 0.00 | $-0.078 \pm 0.020$ | $-0.89 \pm 0.07$ | 8.63 |
| 0.20 | $-0.069 \pm 0.020$ | $-0.93 \pm 0.08$ | 8.57 |
| 0.40 | $-0.058 \pm 0.017$ | $-0.99 \pm 0.06$ | 8.48 |
| 0.59 | $-0.049 \pm 0.015$ | $-1.03 \pm 0.06$ | 8.40 |
| 0.79 | $-0.040 \pm 0.014$ | $-1.08 \pm 0.05$ | 8.28 |
| 1.02 | $-0.034 \pm 0.012$ | $-1.12 \pm 0.05$ | 8.16 |
| 1.21 | $-0.031 \pm 0.011$ | $-1.13 \pm 0.04$ | 8.13 |
| 1.42 | $-0.020 \pm 0.008$ | $-1.18 \pm 0.03$ | 7.94 |
| 1.59 | $-0.014 \pm 0.008$ | $-1.21 \pm 0.03$ | 7.81 |
| 1.79 | $-0.017 \pm 0.005$ | $-1.20 \pm 0.02$ | 7.86 |
| 2.02 | $-0.012 \pm 0.006$ | $-1.23 \pm 0.02$ | 7.71 |
| 2.24 | $-0.008 \pm 0.005$ | $-1.25 \pm 0.02$ | 7.61 |
| 2.40 | $-0.007 \pm 0.005$ | $-1.25 \pm 0.02$ | 7.61 |
| 2.59 | $-0.009 \pm 0.003$ | $-1.25 \pm 0.01$ | 7.61 |
| 2.80 | $-0.003 \pm 0.002$ | $-1.28 \pm 0.01$ | 7.43 |
| 3.04 | $0.001 \pm 0.003$ | $-1.30 \pm 0.01$ | 7.31 |
| 3.18 | $-0.002 \pm 0.002$ | $-1.28 \pm 0.00$ | 7.43 |
| 3.48 | $0.000 \pm 0.004$ | $-1.30 \pm 0.01$ | 7.31 |
| 3.66 | $0.000 \pm 0.003$ | $-1.30 \pm 0.01$ | 7.31 |
| 3.86 | $0.000 \pm 0.003$ | $-1.30 \pm 0.01$ | 7.31 |
| 4.07 | $0.004 \pm 0.001$ | $-1.32 \pm 0.01$ | 7.17 |
| 4.31 | $0.004 \pm 0.001$ | $-1.33 \pm 0.01$ | 7.10 |
| 4.59 | $0.002 \pm 0.002$ | $-1.32 \pm 0.01$ | 7.17 |
| 4.89 | $0.001 \pm 0.001$ | $-1.32 \pm 0.01$ | 7.17 |
| 5.25 | $0.008 \pm 0.002$ | $-1.35 \pm 0.01$ | 6.95 |
| 5.66 | $0.003 \pm 0.003$ | $-1.34 \pm 0.01$ | 7.03 |
| 6.14 | $0.010 \pm 0.000$ | $-1.37 \pm 0.01$ | 6.79 |
| 6.73 | $0.005 \pm 0.003$ | $-1.36 \pm 0.01$ | 6.87 |
| 7.44 | $0.007 \pm 0.004$ | $-1.37 \pm 0.01$ | 6.79 |
| 8.35 | $0.006 \pm 0.002$ | $-1.38 \pm 0.01$ | 6.70 |
| 9.55 | $0.005 \pm 0.002$ | $-1.39 \pm 0.01$ | 6.61 |
| 11.21 | $0.002 \pm 0.002$ | $-1.40 \pm 0.01$ | 6.52 |

**Table A4.**

| Redshift | $grad$ (dex kpc$^{-1}$) | M 33 $\log(N/O)_0$ | $12+\log(O/H)_0$ |
|---|---|---|---|
| 0.00 | $-0.080 \pm 0.023$ | $-0.93 \pm 0.07$ | 8.62 |
| 0.20 | $-0.070 \pm 0.022$ | $-0.97 \pm 0.07$ | 8.56 |
| 0.40 | $-0.056 \pm 0.019$ | $-1.04 \pm 0.06$ | 8.40 |
| 0.60 | $-0.045 \pm 0.016$ | $-1.08 \pm 0.05$ | 8.24 |
| 0.80 | $-0.033 \pm 0.013$ | $-1.13 \pm 0.04$ | 8.05 |
| 1.00 | $-0.028 \pm 0.011$ | $-1.16 \pm 0.03$ | 7.94 |
| 1.20 | $-0.023 \pm 0.012$ | $-1.18 \pm 0.04$ | 7.84 |
| 1.40 | $-0.014 \pm 0.009$ | $-1.22 \pm 0.03$ | 7.64 |
| 1.60 | $-0.012 \pm 0.007$ | $-1.23 \pm 0.02$ | 7.60 |
| 1.80 | $-0.011 \pm 0.005$ | $-1.23 \pm 0.01$ | 7.58 |
| 2.00 | $-0.009 \pm 0.004$ | $-1.25 \pm 0.01$ | 7.47 |
| 2.20 | $-0.003 \pm 0.005$ | $-1.27 \pm 0.01$ | 7.35 |
| 2.40 | $-0.001 \pm 0.006$ | $-1.28 \pm 0.02$ | 7.28 |
| 2.60 | $+0.000 \pm 0.004$ | $-1.29 \pm 0.01$ | 7.27 |
| 2.80 | $+0.001 \pm 0.001$ | $-1.30 \pm 0.00$ | 7.18 |
| 3.00 | $+0.005 \pm 0.003$ | $-1.31 \pm 0.01$ | 7.14 |
| 3.20 | $+0.001 \pm 0.001$ | $-1.30 \pm 0.00$ | 7.16 |
| 3.50 | $+0.002 \pm 0.003$ | $-1.31 \pm 0.01$ | 7.13 |
| 3.80 | $-0.000 \pm 0.004$ | $-1.30 \pm 0.01$ | 7.18 |
| 4.00 | $+0.004 \pm 0.003$ | $-1.32 \pm 0.01$ | 7.04 |
| 4.30 | $+0.001 \pm 0.004$ | $-1.31 \pm 0.01$ | 7.13 |
| 4.60 | $+0.010 \pm 0.002$ | $-1.35 \pm 0.01$ | 6.86 |
| 4.90 | $+0.002 \pm 0.001$ | $-1.32 \pm 0.01$ | 7.02 |
| 5.20 | $+0.005 \pm 0.003$ | $-1.33 \pm 0.01$ | 6.99 |
| 5.60 | $-0.001 \pm 0.001$ | $-1.32 \pm 0.01$ | 7.04 |
| 6.10 | $+0.008 \pm 0.002$ | $-1.35 \pm 0.01$ | 6.84 |
| 6.70 | $+0.003 \pm 0.006$ | $-1.35 \pm 0.02$ | 6.88 |
| 7.40 | $+0.005 \pm 0.006$ | $-1.36 \pm 0.02$ | 6.80 |
| 8.30 | $+0.002 \pm 0.002$ | $-1.36 \pm 0.01$ | 6.76 |
| 9.50 | $+0.002 \pm 0.002$ | $-1.38 \pm 0.01$ | 6.64 |
| 11.20 | $+0.004 \pm 0.002$ | $-1.40 \pm 0.01$ | 6.46 |